\documentclass[twocolumn,10pt]{IEEEtran}

\usepackage{amsmath}
\usepackage{mathtools}
\usepackage{enumerate}
\usepackage{paralist}
\usepackage{graphicx}
\usepackage{epstopdf}
\graphicspath{{Figures/}}
\usepackage{subfigure}
\usepackage{cite}
\usepackage{array}
\usepackage{booktabs}
\usepackage{comment}
\usepackage{cases}
\usepackage{algorithm}
\usepackage{algorithmic}
\usepackage{mathrsfs} 
\usepackage{url}
\usepackage{amssymb}
\usepackage{amsthm} 

\usepackage{bigints} 

\usepackage[table]{xcolor}

\usepackage{ctable}

\usepackage{array}
\usepackage{ragged2e}
\newcolumntype{P}[1]{>{\RaggedRight\hspace{0pt}}p{#1}}
\newcolumntype{C}[1]{>{\centering\hspace{0pt}}p{#1}}

\newtheorem{theorem}{Theorem}

\newtheorem{remark}{Remark}

\newtheorem{appxlem}{Lemma}[section] 

\begin{document}

\title{Distributed Resource Allocation for Relay-Aided Device-to-Device Communication: A Message Passing Approach}
\author{Monowar Hasan and Ekram Hossain
\thanks{M. Hasan and E. Hossain are with the Department of Electrical and Computer Engineering, University of Manitoba, Winnipeg, Canada (emails:  monowar\_hasan@umanitoba.ca, Ekram.Hossain@umanitoba.ca).}} 

\maketitle

\begin{abstract}

Device-to-device (D2D) communication underlaying cellular wireless networks is a promising concept to improve user experience and resource utilization by allowing direct transmission between two cellular devices.  In this paper, performance of network-assisted D2D communication is investigated where D2D traffic is carried through relay nodes. Considering a multi-user and multi-relay network, we propose a distributed solution for resource allocation with a view to maximizing network sum-rate. An optimization problem is formulated for  radio resource allocation at the relays. The objective is to maximize end-to-end rate as well as satisfy the data rate requirements for cellular and D2D user equipments under total power constraint. Due to intractability of the resource allocation problem, we propose a solution approach using message passing technique  where each user equipment sends and receives information messages to/from the relay node in an iterative manner with the goal of achieving an optimal allocation. Therefore, the computational effort is distributed among all the user equipments and the corresponding relay node. The convergence and optimality of the proposed scheme are proved and a possible distributed implementation of the scheme in practical LTE-Advanced networks is outlined. The numerical results show that there is a distance threshold beyond which relay-aided D2D communication significantly improves network performance with a small increase in end-to-end delay when compared to direct communication between D2D peers.
\end{abstract}


\section{Introduction} \label{sec:intro}

In recent years, new applications such as content distribution and location-aware advertisement underlaying cellular networks have drawn much attention to end-users and network providers. The emergence of such new applications brings D2D communication under intensive discussions in  academia, industry, and standardization bodies. The concept of D2D communication has been introduced to allow local peer-to-peer transmission among user equipments (UEs) bypassing the base station [e.g., eNB in a Long Term Evolution Advanced (LTE-A) network] to cope with high data rate services (i.e., video sharing, online gaming, proximity-aware social networking). D2D communication was first
proposed in \cite{d2d_first_relay} to enable multi-hop relaying in cellular networks. In addition to traditional local voice and data services, other potential D2D use-cases have been introduced in the literature such as peer-to-peer communication, local advertisement, multi-player gaming, data flooding \cite{d2d_example, d2d_example2, 3gpp:d2d_example}, multicasting \cite{d2d_example_multicast}, \cite{d2d_multicast}, video dissemination \cite{d2d_example_video, d2d_example_video2, d2d_inceremental_relay}, and machine-to-machine (M2M) communication \cite{d2d_m2m_1}.

Using local data transmissions, D2D communication offers the following advantages: \begin{inparaenum}[\itshape i\upshape)]  
\item extended coverage;
\item offloading users from cellular networks \cite{d2d_example_offload};
\item increased throughput and spectrum efficiency as well as improved energy efficiency \cite{d2d_energy}.
\end{inparaenum}
However, in a D2D-enabled network, a number of practical considerations may limit the advantages of D2D communication. In practice, setting up reliable direct links between the D2D UEs while satisfying the quality-of-service (QoS) requirements of both the traditional cellular UEs (CUEs) as well as the  D2D UEs is challenging due to the following reasons:  

\begin{enumerate} [\itshape i\upshape)]
\item \textit{Large distance:} the potential D2D UEs may not be in near proximity;
\item \textit{Poor propagation condition:}  the link quality between potential D2D UEs may not be favorable for direct communication;
\item \textit{Interference to and from CUEs:} in an underlay system, without an efficient power control mechanism, the D2D transmitters may cause severe interference to other receiving nodes. The D2D receivers may also experience interference from CUEs and/or eNB. One remedy to this problem is to partition the available spectrum (i.e., use overlay D2D communication). However, this can significantly reduce the spectrum utilization.
\end{enumerate} 

Network-assisted transmissions through relays could efficiently enhance the performance of D2D communication when the D2D UEs are too far away from each other or the quality of the channel between the UEs is not good enough for direct communication. Unlike the existing literature on D2D communication, in this paper, we consider relay-assisted D2D communication
underlaying LTE-A cellular networks where D2D UEs are served by the relay nodes. We utilize the self-backhauling configuration of LTE-A Layer-3 (L3) relay which enables it to perform operations similar to those of an eNB. We consider scenarios in which the potential D2D UEs are located in the same macrocell (i.e., office blocks or university areas, concert halls etc.); however, the proximity and link condition may not be favorable for direct communication. Therefore, they communicate via relays. 
The radio resources  (e.g., resource blocks [RBs] and power) at the relays are shared among the D2D communication links and the two-hop cellular links using these relays. 

The goal of this work is to design a practical resource allocation algorithm for network-assisted D2D communications. We show that the resource allocation problem can be converted to a \textit{max-sum message passing} (MP) problem over a graphical model. The MP algorithms have been recognized as powerful tools that can be used to solve many problems in signal processing, coding theory, machine learning, natural language processing and computer vision. When MP is applied to solve a problem, the messages represent probabilities (i.e., beliefs) exchanged with the goal of achieving optimal decisions. Analogously, in the context of the resource allocation problem  for relay-aided D2D communication, the MP strategy can be applied to pass messages between UEs and relays until a global allocation is obtained. The advantage of applying MP strategy in resource allocation is that it provides a low-complexity distributed solution and reduces the computation burden at the controller node. Motivated by the above fact, in this work, we apply the max-sum variation of the message passing technique to represent the resource allocation problem by a \textit{factor graph}. To this end, we propose a distributed solution approach with polynomial time-complexity and low signaling overhead. The main contributions of this paper can be summarized as follows:

\begin{itemize}

\item We model and analyze the performance of relay-assisted D2D communication. The problem of RB and power allocation at the relay nodes for the CUEs and D2D UEs is formulated. As opposed to most of the resource allocation schemes in the literature where only a single D2D link is considered, we consider multiple D2D links along with
multiple cellular links that are supported by the relay nodes.

\item We provide a novel solution technique using message passing. 
Utilizing message passing strategy, we provide a low-complexity distributed solution by which resource blocks and transmission power can be allocated in a distributed fashion.

\item We analyze the complexity and the optimality of the solution. To this end, we compare the performance of our relay-based D2D communication scheme with a direct D2D communication method and observe that relaying improves network performance for distant D2D peers without increasing the end-to-end delay significantly.
\end{itemize}

The remainder of this paper is organized as follows. A review of
the existing work in the literature and motivation behind this work are presented in Section \ref{sec:related_works}. Followed by the system model and related assumptions in Section \ref{sec:sys_model}, we formulate the resource allocation problem in Section \ref{sec:RAP}. The message passing strategy to solve the resource allocation problem is introduced in Section \ref{sec:rap_mp} and a distributed solution is proposed in Section \ref{sec:distributed_sol}. The performance evaluation  results are presented in Section \ref{sec:performance}.  We conclude the paper in Section \ref{sec:conclusion}. The key mathematical notations used in the paper are listed in Table \ref{tab:notations_mp}.

\begin{table}[!t]
\renewcommand{\arraystretch}{1.3}
\caption{Mathematical Notations}
\label{tab:notations_mp}
\centering
\begin{tabular}{c|p{5.2cm}}
\hline
\bfseries Notation & \bfseries Physical interpretation\\
\hline\hline
$\mathcal{N} = \lbrace 1, 2, \ldots, N \rbrace $ & Set of available RBs \\
\hline $\mathcal{L} = \lbrace 1, 2, \ldots, L \rbrace$ & Set of relays \\
\hline $\mathcal{C} = \lbrace 1, 2, \ldots, C \rbrace $, & Set of CUEs and D2D UEs, respectively \\
$\mathcal{D} = \lbrace 1, 2, \ldots, D \rbrace $ & \\
\hline $\mathcal{U}_l, |\mathcal{U}_l|$ & Set of UEs and total number of UEs served by relay $l$, respectively \\
\hline $u_l$  & A UE served by relay $l$ \\
\hline $\gamma_{u_l, l, 1}^{(n)}, \gamma_{l, u_l, 2}^{(n)}$ & ${\rm SINR}$ for UE $u_l$ over RB $n$ is first and second hop, respectively \\
\hline $R_{u_l}^{(n)}$ & End-to-end data rate for $u_l$ over RB $n$ \\
\hline $Q_{u_l}$ & Data rate requirement for $u_l$\\
\hline $x_{u_l}^{(n)}, P_{u_l, l}^{(n)}$ & RB allocation indicator and actual transmit power for $u_l$ over RB $n$, respectively \\
\hline $\mathbf{x}_{\boldsymbol l}, \mathbf{P}_{\boldsymbol l}$ & RB and power allocation vector\\
\hline $\mathfrak{R}_{n,l}(\cdot), \mathfrak{W}_{u_l, l}(\cdot)$ & Utility functions in factor graph dealing with optimization constraints \\
\hline $\delta_{\mathfrak{R}_{n,l}(\cdot) \rightarrow x_{u_l}^{(n)}} \left( x_{u_l}^{(n)} \right)$ & Message from function node $\mathfrak{R}_{n,l}(\cdot)$ to variable node $x_{u_l}^{(n)}$ \\
\hline $\delta_{\mathfrak{W}_{u_l, l}(\cdot)  \rightarrow x_{u_l}^{(n)}} \left( x_{u_l}^{(n)} \right)$ & Message from $\mathfrak{W}_{u_l, l}(\cdot)$ function node to any variable node $x_{u_l}^{(n)}$  \\
\hline $\phi_{u_l}^{(n)}\left( x_{u_l}^{(n)} \right)$ & Marginal at variable node $x_{u_l}^{(n)}$ in factor graph \\
\hline $\psi_{u_l, l}^{(n)}, \tilde{\psi}_{u_l, l}^{(n)}$  & Normalize messages for UE $u_l$ over RB $n$ \\
\hline $\langle \upsilon_{u_l}^{(j)}\rangle_{z \setminus n}$ & $z$-th sorted element of $\boldsymbol {\chi_{u_l}}$ without considering the term $R_{u_l}^{(n)} + \tilde{\psi}_{u_l, l}^{(n)}$ \\
\hline $\tau_{u_l, l}^{(n)} $ & Node marginal for UE $u_l$ over RB $n$ \\
\hline $\kappa_{u_l}$ & Required number RB(s) for $u_l$ to satisfy the rate requirement $Q_{u_l}$\\
\hline $\mathbf{abs}\lbrace y \rbrace$ & Absolute value of variable $y$ \\
\hline $\mathfrak{D}_{\rm 2 hop}$ & End-to-end delay for two hop relay-aided communication \\
\hline 
\end{tabular}
\end{table}

\section{Related Work and Motivation} \label{sec:related_works}

Although resource allocation for D2D communication in orthogonal frequency-division multiple access (OFDMA)-based wireless networks is one of the active areas of research, only a very few work in the literature consider relays for D2D communication. A summary of related literature and comparison with our proposed scheme is presented in Table \ref{tab:summary}.

\begin{table*}[!t]
\renewcommand{\arraystretch}{1.3}
\caption{Summary of Related Work and Proposed Scheme}
\label{tab:summary}
\centering
\begin{tabular}{C{1.7cm}| P{2.7cm}| l| P{3.3cm}| p{2.0cm}| p{2.0cm} }
\hline
\bfseries Reference & \bfseries Problem focus & \bfseries Relay-aided & \bfseries Solution approach & \bfseries Solution type & \bfseries Optimality\\
\hline\hline
\cite{d2d_multicast} & Theoretical analysis, spectrum utilization & No & Iterative cluster partitioning & Centralized & Optimal \\
\cite{d2d_inceremental_relay} & Resource allocation & No & Proposed heuristic & Centralized & Suboptimal\\
\cite{zul-d2d} & Resource allocation & No & Proposed greedy heuristic & Centralized & Suboptimal\\
\cite{d2d_new_paper} & Resource allocation & No\textsuperscript{*} & Numerical optimization & Semi-distributed & Pareto optimal \\
 \cite{d2d_swarm} & Resource allocation, mode selection & No & Particle swarm optimization & Centralized & Suboptimal \\  
\cite{d2d_intf_graph} & Resource allocation & No & Interference graph coloring & Centralized & Suboptimal \\
\cite{phond-d2d} & Resource allocation & No & Column generation based greedy heuristic & Centralized & Suboptimal \\
\cite{le_d2d} & Resource allocation & No & Two-phase heuristic & Centralized & Suboptimal \\
\cite{d2d-rel-1} & Theoretical analysis, performance evaluation  & Yes & Statistical analysis  & Centralized & Optimal\\
\cite{d2d_relay_2} & Performance evaluation  & Yes & Heuristic, simulation  & Centralized & N/A\textsuperscript{\textdagger} \\
\cite{d2d_our_paper} & Resource allocation & Yes & Numerical optimization & Centralized & Asymptotically optimal \\
\hline
\textit{Proposed scheme} & Resource allocation & Yes & Max-sum message passing & Distributed & Asymptotically optimal \\
\hline
\multicolumn{6}{l}{\textsuperscript{*}\footnotesize{D2D UEs serve as relays to assist CUE-eNB communications.}} \\
\multicolumn{6}{l}{\textsuperscript{\textdagger}\footnotesize{No information is available.}} \\
\end{tabular}
\end{table*}

In \cite{zul-d2d}, a greedy heuristic-based resource allocation scheme is proposed for both uplink and downlink scenarios where a D2D pair shares the same resources with CUE only if the achieved ${\rm SINR}$ is greater than a given ${\rm SINR}$ requirement. A new spectrum sharing protocol for D2D communication overlaying a cellular network is proposed in \cite{d2d_new_paper}, which allows the D2D users to communicate bi-directionally while assisting the two-way communications between the eNB and the CUE. 
In \cite{d2d_swarm}, the mode selection and resource allocation problem for D2D communication underlaying cellular networks is investigated and the solution is obtained by particle swarm optimization. Through simulations, the authors show that the proposed scheme improves system performance compared to overlay D2D communication. In \cite{d2d_multicast}, D2D communication is proposed to improve the performance of multicast transmission among the members of a multicast group.  A graph-based resource allocation method for cellular networks with underlay D2D communication is proposed in \cite{d2d_intf_graph}. Due to the intractability of resource allocation problem, the authors propose a sub-optimal graph-based approach which accounts for interference and capacity of the network. A resource allocation scheme based on a column generation method is proposed in \cite{phond-d2d} to maximize the spectrum utilization by finding the minimum transmission length (i.e., time slots) for D2D links while protecting the cellular users from interference and guaranteeing QoS. 
A two-phase resource allocation scheme for cellular
network with underlaying D2D communication is proposed in \cite{le_d2d}. Due to $\mathcal{NP}$-hardness of the optimal resource allocation problem, the author proposes a two-phase low-complexity sub-optimal solution where after performing optimal resource allocation for cellular users, a heuristic subchannel allocation scheme for D2D flows is applied which initiates the resource allocation from the flow with the minimum rate requirements. 

In \cite{d2d_inceremental_relay}, the authors  propose an incremental relay mode for D2D communication where D2D transmitters multicast to both the D2D receiver and BS. In case the D2D transmission fails, the BS retransmits
the multicast message to the D2D receiver. Although the base station receives a copy of the D2D message which is retransmitted in case of failure,  this incremental relay mode of communication consumes part of the downlink resources for retransmission and reduces spectrum utilization. In \cite{d2d-rel-1,  d2d_relay_2}, the maximum ergodic capacity and outage probability of cooperative relaying is investigated in relay-assisted D2D communication  considering power constraints at the eNB. The numerical results show that multi-hop relaying lowers the outage probability and improves cell edge capacity by reducing the effect of interference from the CUE. It is worth noting that most of the above cited works provide centralize solutions. Besides, in \cite{zul-d2d, d2d_new_paper, phond-d2d, d2d_swarm, d2d_multicast, d2d_intf_graph, le_d2d, d2d_inceremental_relay}, the effect of using relays in D2D communication is not studied. As a matter of fact, relaying mechanism explicitly in context of D2D communication has not been studied comprehensively in the literature. 

Taking the advantage of L3 relays supported by the 3GPP standard, in our earlier work \cite{d2d_our_paper}, we studied the performance of network-assisted D2D communication and showed that relay-aided D2D communication provides significant performance gain for long distance D2D links. However, the proposed solution in \cite{d2d_our_paper} is obtained in a centralized manner by a central controller (i.e., L3 relay). In this work, we develop a \textit{distributed solution} technique utilizing the MP strategy on a factor graph. Factor graph and other graphical modes have been used as powerful solution techniques to tackle a wide range of problems in various domains; however, they have not been commonly used in the context of resource allocation in cellular wireless networks. 

To the best of our knowledge, the MP scheme for resource allocation in wireless networks was first introduced in \cite{mp-icc-09} to minimize the transmission power in the uplink of a multi-carrier system. A resource allocation scheme based on MP is proposed in \cite{mp-dft} for DFT-Spread-OFDMA uplink communication. In \cite{mp-twireless}, the message passing approach is used to allocate resources to minimize the transmission power for both single and multiple transmission formats in an OFDMA-based cellular network. In \cite{cr-pgm-jsac}, a message passing algorithm is proposed for a cognitive radio network to find assignment of secondary users to detect primary users so that the best overall network performance is achieved in a computationally efficient manner. Different from the above works, to allocate radio resource efficiently in a relay-aided D2D communication scenario, we use the max-sum MP strategy in our problem domain and propose a distributed solution in order to maximize the spectrum utilization. To this end, we analyze the complexity of the proposed solution and prove its optimality and convergence. We also discuss the delay performance and present an approach for possible implementation of our proposed solution in the LTE-A network setup. 

\section{System Model and Assumptions} \label{sec:sys_model}

\subsection{Network Model}

\begin{figure}[!t]
\centering
\includegraphics[scale=0.60]{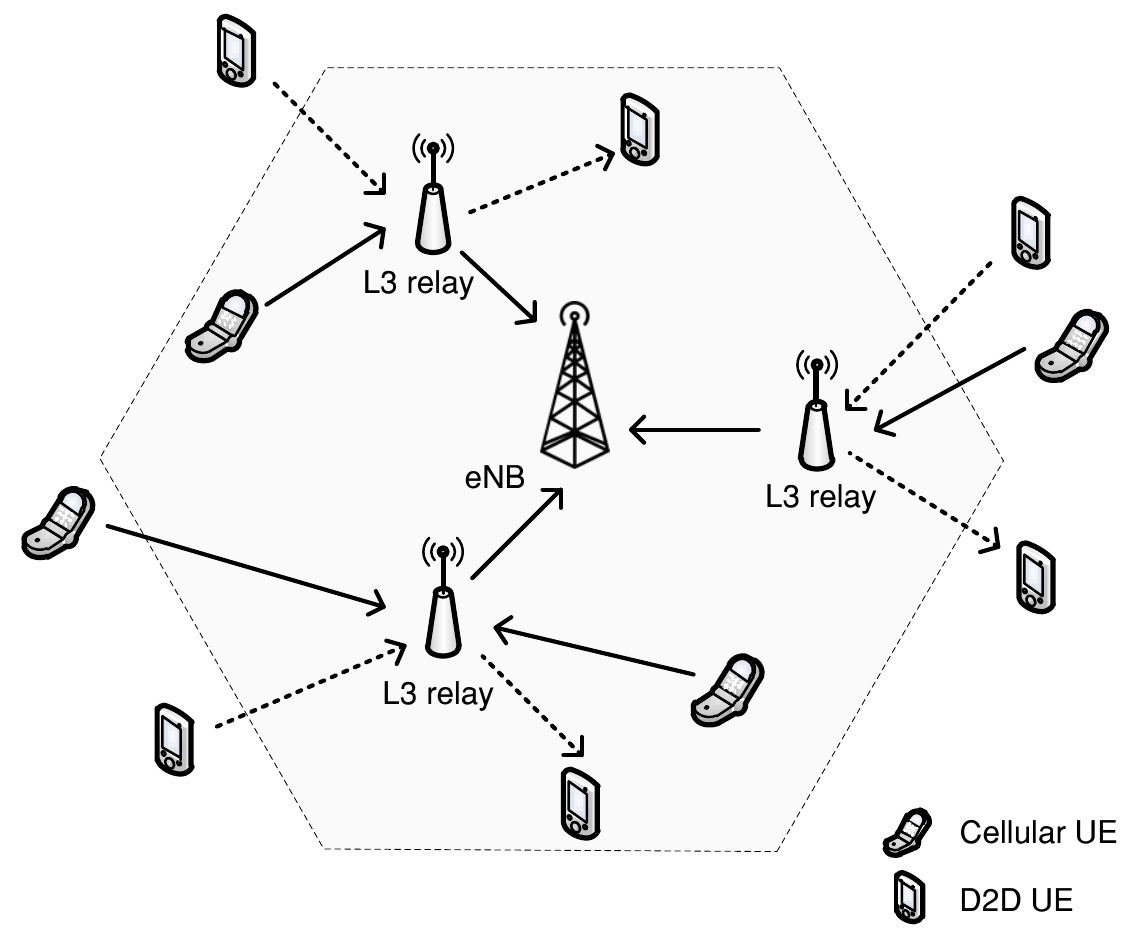}
\caption{A single cell with multiple relay nodes. We assume that the CUE-eNB links are unfavorable for direct communication and need the assistance of relays. The D2D UEs are also supported by the relay nodes due to long distance and/or poor link condition between peers. 
} 
\label{fig:nw_diagram}
\end{figure}

Let us consider a D2D-enabled  cellular network with multiple relays as shown in Fig. \ref{fig:nw_diagram}. A relay node in LTE-A is connected to the radio access network (RAN) through a donor eNB with a wireless connection and serves  both the cellular UEs and D2D UEs. Let $\mathcal{L} = \lbrace 1, 2, \ldots, L \rbrace$ denote the set of fixed-location Layer 3 (L3) relays\footnote{An L3 relay with self-backhauling configuration performs the same operation as an eNB except that it has a lower transmit power and a smaller cell size. It controls cell(s) and each cell has its own cell identity. The relay transmits its own control signals and the UEs receive scheduling information directly from the relay node \cite{relay-book-1}.} in the network. The system bandwidth is divided into $N$ RBs denoted by $\mathcal{N} = \lbrace 1, 2, \ldots, N \rbrace$. When the link condition between two D2D UEs is too poor for direct communication,  scheduling and resource allocation for the D2D UEs can be done in a relay node (i.e., L3 relay) and the D2D traffic can be transmitted through that relay. We refer to this as \textit{relay-aided D2D communication} which can be an efficient approach to provide a better QoS  (e.g., data rate) for communication between distant D2D UEs.

The CUEs and D2D UEs constitute set $\mathcal{C} = \lbrace 1, 2, \ldots, C \rbrace$ and $\mathcal{D} = \lbrace 1, 2, \ldots, D \rbrace$, respectively, where the pairs of D2D UEs are discovered during the D2D session setup. We assume that the CUEs are outside the coverage region of eNB and/or having bad channel condition, and therefore, CUE-eNB communications need to be supported by the relays. Besides, the direct communication between two D2D UEs  requires the assistance of a relay node. The UEs (i.e., both cellular and D2D UEs) assisted by relay $l$ are denoted by $u_l$. The set of UEs assisted by relay $l$ is $\mathcal{U}_l$ such that $\mathcal{U}_l \subseteq \lbrace \mathcal{C} \cup \mathcal{D} \rbrace, \forall l \in \mathcal{L}$, $\bigcup_l \mathcal{U}_l = \lbrace \mathcal{C} \cup \mathcal{D} \rbrace$, and  $\bigcap_l \mathcal{U}_l = \varnothing$.  In the second hop, there could be multiple relays transmitting to their associated D2D UEs. We assume that the relays transmit to the eNB using orthogonal channels and this scheduling of relays is done by the eNB\footnote{Scheduling of relay nodes by the eNB is out of the scope of this paper.}. According to our system model, taking the advantage of L3 relays, scheduling and resource allocation for the UEs is performed in the relay node to reduce the computational load at the eNB.

\subsection{Radio Propagation Model}

For modeling the  propagation channel, we consider distance-dependent path-loss and shadow fading; furthermore, the channel is assumed to  experience Rayleigh fading. In particular, we consider realistic 3GPP propagation environment\footnote{Any other propagation model for D2D communication can be used for the proposed resource allocation method.} presented in \cite{relay-book-2}. For example, UE-relay (and relay-D2D) link follows the following path-loss equation: 
\begin{equation}
PL_{u_l,l}(\ell)_{[dB]} = 103.8 + 20.9 \log(\ell) + L_{su} + 10 \log(\varsigma)
\end{equation}
where $\ell$ is the distance between UE and relay in kilometer, $L_{su}$ accounts for shadow fading and is modelled as a log-normal random variable, and $\varsigma$ is an exponentially distributed random variable which represents the Rayleigh fading channel power gain. Similarly, the path-loss equation for relay-eNB link is expressed as \begin{equation}
PL_{l,eNB}(\ell)_{[dB]} = 100.7 + 23.5 \log(\ell) + L_{sr} + 10 \log(\varsigma)
\end{equation} 
where $L_{sr}$ is a log-normal random variable accounting for shadow fading. Hence given the distance $\ell$, the link gain between any pair of network nodes $i, j$ can be calculated as $10^{-\frac{PL_{i,j}(\ell)}{10}}$.

\subsection{Achievable Data Rate}

We denote by $h_{i,j}^{(n)}$ the direct link gain between node $i$ and $j$ over RB $n$. The interference link gain between relay (UE) $i$  and UE (relay) $j$ over RB $n$ is denoted by $g_{i,j}^{(n)}$  where UE (relay) $j$ is not associated with relay (UE) $i$.  The unit power ${\rm SINR}$ for the link between UE $u_l \in \mathcal{U}_l$ and relay $l$ using RB $n$ in the first hop is given by
\begin{equation}
\label{eq:SINR_1}
\gamma_{u_l, l, 1}^{(n)} = \frac{h_{u_l, l}^{(n)}}{\displaystyle \sum_{\substack{\forall u_j \in \mathcal U_j, \\ j \neq l, j \in \mathcal{L}}} P_{u_j, j}^{(n)} g_{u_j, l}^{(n)} + \sigma^2}.
\end{equation}
The unit power ${\rm SINR}$ for the link between relay $l$ and eNB  for CUE $u_l$ (i.e.,  $u_l \in \lbrace \mathcal{C} \cap \mathcal{U}_l \rbrace$) in the second hop is as follows: 
\begin{equation}
\label{eq:SINR_2}
\gamma_{l, u_l, 2}^{(n)} = \frac{h_{l, eNB}^{(n)}}{\displaystyle \sum_{\substack{\forall u_j \in \lbrace \mathcal{D} \cap \mathcal{U}_j \rbrace, \\ j \neq l, j \in \mathcal{L}}} P_{j, u_j}^{(n)} g_{j, eNB}^{(n)} + \sigma^2}.
\end{equation}
Similarly, the unit power ${\rm SINR}$ for the link between relay $l$ and receiving D2D UE for the D2D UEs $u_l$ (i.e., $u_l \in \lbrace \mathcal{D} \cap \mathcal{U}_l \rbrace$) in the second hop can be written as
\begin{equation}
\label{eq:SINR_3}
\gamma_{l, u_l, 2}^{(n)} = \frac{h_{l, u_l}^{(n)}}{\displaystyle \sum_{\substack{\forall u_j \in  \mathcal{U}_j, \\ j \neq l, j \in \mathcal{L}}} P_{j, u_j}^{(n)} g_{j, u_l}^{(n)} + \sigma^2}.
\end{equation}

In  (\ref{eq:SINR_1})--(\ref{eq:SINR_3}), $P_{i, j}^{(n)}$ is the transmit power in the  link between $i$ and $j$ over RB $n$, $\sigma^2 =  N_0 B_{RB}$, where $B_{RB}$ is bandwidth of an RB, and $N_0$ denotes thermal noise.  $h_{l, eNB}^{(n)}$ is the gain in the relay-eNB link and $h_{l, u_l}^{(n)}$ is the gain in the link between relay $l$ and receiving D2D UE corresponding to the D2D transmitter UE $u_l$.

The achievable data rate\footnote{If there is no relay in the network, the achievable data rate for the UE $u$ over RB $n$ can be expressed as $\widetilde{r}_{u}^{(n)} = B_{RB} \log_2 \left( 1 + P_{u}^{(n)} \tilde{\gamma}_{u}^{(n)} \right)$, where $\tilde{\gamma}_{u}^{(n)}  = \frac{h_{u, \hat{u}}^{(n)}}{\displaystyle \sum_{\forall j \in \hat{{\mathcal U}}_u} P_{j}^{(n)} g_{u, j}^{(n)} + \sigma^2} $, $h_{u, \hat{u}}^{(n)}$ is the channel gain between CUE-eNB link ($u \in \mathcal{C}$) or the channel gain between D2D UEs ($u \in \mathcal{D}$) and  $\hat{{\mathcal U}}_u$ denotes the set of UEs transmitting with same RB(s) as $u$.} for $u_l$ in the first hop can be expressed as $$r_{u_l, 1}^{(n)} = B_{RB} \log_2 \left( 1 + P_{u_l, l}^{(n)} \gamma_{u_l, l, 1}^{(n)} \right).$$ Similarly, the achievable data rate in the second hop is given by $$r_{u_l, 2}^{(n)} = 
B_{RB} \log_2 \left( 1 + P_{l, u_l}^{(n)} \gamma_{l, u_l, 2}^{(n)} \right).$$ Since we are considering a two hop communication approach, the end-to-end data rate for  $u_l$ on RB $n$ is the half of minimum achievable data rate over two hops, i.e.,   
\begin{equation}
\label{eqn:e2e_rate}
R_{u_l}^{(n)} =  \frac{1}{2} \min \left\lbrace  r_{u_l, 1}^{(n)} ,~ r_{u_l, 2}^{(n)}    \right\rbrace.
\end{equation}

\section{Formulation of the Resource Allocation Problem} 
\label{sec:RAP}

For each relay, the objective of resource (i.e., RB and transmit power) allocation problem (RAP) is to obtain  the assignment of RB and power level to the UEs that maximizes the system capacity, which is defined as the minimum achievable data rate over two hops. The RB allocation indicator is denoted by a binary decision variable $x_{u_l}^{(n)} \in \lbrace 0, 1\rbrace$, where 
\begin{equation}
x_{u_l}^{(n)}  = \begin{cases}
 1, \quad  \text{if RB $n$ is assigned to UE $u_l$} \\
 0, \quad \text{otherwise.}
\end{cases}
\end{equation}

Hence, the objective of RAP is to obtain the RB and power allocation vectors for each relay $l \in \mathcal{L}$, i.e., 
$\mathbf{x}_{\boldsymbol l} = \left[ x_{1}^{(1)}, \ldots,  x_{1}^{(N)}, \ldots, x_{|\mathcal{U}_l|}^{(1)}, \ldots, x_{|\mathcal{U}_l|}^{(N)} \right]^\mathsf{T}$ and $\mathbf{P}_{\boldsymbol l} = \left[ P_{1,l}^{(1)}, \ldots, P_{1,l}^{(N)}, \ldots,  P_{|\mathcal{U}_l|,l}^{(1)}, \ldots, P_{|\mathcal{U}_l|,l}^{(N)} \right]^\mathsf{T}$ respectively, which maximize the data rate. Let the maximum allowable transmit power for UE (relay) is $P_{u_l}^{max}$ ($P_l^{max}$). Let the QoS (i.e., data rate) requirements for UE $u_l$ is denoted by $Q_{u_l}$ and $\displaystyle R_{u_l} = \sum_{n =1}^N  x_{u_l}^{(n)} R_{u_l}^{(n)}$ denotes the achievable sum-rate over allocated RB(s). Considering that the same RB(s) will be used by the relay in both the hops, the resource allocation problem for each relay $l \in \mathcal{L}$ can be stated as follows:
\begin{subequations}
\setlength{\arraycolsep}{0.0em}
\begin{eqnarray}
\mathbf{(P1)} ~ \underset{x_{u_l}^{(n)}, P_{u_l, l}^{(n)}, P_{l, u_l}^{(n)}}{\operatorname{max}} ~ \sum_{u_l \in \mathcal{U}_l } &&  \sum_{n =1}^N   x_{u_l}^{(n)} R_{u_l}^{(n)}  \nonumber \\
\text{subject to} \quad \sum_{u_l \in \mathcal{U}_l} x_{u_l}^{(n)} ~ &&{\leq} ~ 1,  \quad ~~~\forall n \in \mathcal{N} \label{eq:con-bin} \\
\quad \sum_{n =1}^N x_{u_l}^{(n)} P_{u_l, l}^{(n)} ~ &&{\leq} ~ P_{u_l}^{max}, ~ \forall u_l \in \mathcal{U}_l \label{eq:con-pow-ue} \\
\quad \sum_{u_l \in \mathcal{U}_l } \sum_{n =1}^N x_{u_l}^{(n)} P_{l, u_l}^{(n)} ~ &&{\leq} ~ P_l^{max}  \label{eq:con-pow-rel} \\
\quad \sum_{u_l \in \mathcal{U}_l } x_{u_l}^{(n)} P_{u_l, l}^{(n)} g_{{u_l^*}, l, 1}^{(n)} ~ &&{\leq} ~ I_{th, 1}^{(n)},  ~~\forall n \in \mathcal{N} \label{eq:con-intf-1}\\
\quad \sum_{u_l \in \mathcal{U}_l } x_{u_l}^{(n)}  P_{l, u_l}^{(n)} g_{l, {u_l^*}, 2}^{(n)} ~ &&{\leq} ~ I_{th, 2}^{(n)}, ~~\forall n \in \mathcal{N} \label{eq:con-intf-2} \\
\quad R_{u_l}  ~ &&{\geq} ~ Q_{u_l}, \quad \forall u_l \in \mathcal{U}_l  \label{eq:con-QoS-cue}\\
\quad P_{u_l, l}^{(n)} ~ \geq ~ 0, ~~ P_{l, u_l}^{(n)} ~ &&{\geq} ~ 0,  \quad ~~~\forall n \in \mathcal{N}, u_l \in \mathcal{U}_l \label{eq:con-pow-0}
\end{eqnarray}
\setlength{\arraycolsep}{5pt}
\end{subequations}
where the rate of $u_l$ over RB $n$ 
$$R_{u_l}^{(n)} = \frac{1}{2} \min \left\lbrace \begin{smallmatrix}
B_{RB} \log_2 \left( 1 + P_{u_l, l}^{(n)} \gamma_{u_l, l, 1}^{(n)} \right), \vspace*{0.5em}\\
B_{RB} \log_2 \left( 1 + P_{l, u_l}^{(n)} \gamma_{l, u_l, 2}^{(n)} \right)   
\end{smallmatrix} \right\rbrace,$$ 
the unit power ${\rm SINR}$  for the first hop, \[\gamma_{u_l, l, 1}^{(n)} = \frac{h_{u_l, l}^{(n)}}{ I_{u_l,l,1}^{(n)} + \sigma^2  },\] and the unit power ${\rm SINR}$ for the second hop, 
\begin{numcases}{\gamma_{l, u_l, 2}^{(n)} = }
\frac{h_{l, eNB}^{(n)}}{I_{l, u_l,2}^{(n)} + \sigma^2}, &   $u_l \in \lbrace \mathcal{C} \cap \mathcal{U}_l \rbrace$ \nonumber \\
\frac{h_{l, u_l}^{(n)}}{I_{l,u_l,2}^{(n)} + \sigma^2}, &   $u_l \in \lbrace \mathcal{D} \cap \mathcal{U}_l \rbrace$. \nonumber
\end{numcases}
In the above, $I_{u_l,l,1}^{(n)}$ and $I_{l, u_l,2}^{(n)}$ denote the interference received by $u_l$ over RB $n$ in the first and second hop, respectively, and are given as follows:
 $$I_{u_l,l,1}^{(n)} = \displaystyle \sum_{\substack{\forall u_j \in \mathcal U_j, \\ j \neq l, j \in \mathcal{L}}}  x_{u_j}^{(n)} P_{u_j, j}^{(n)} g_{u_j, l}^{(n)},$$
\begin{numcases}{I_{l,u_l,2}^{(n)} = }
\displaystyle \sum_{\substack{\forall u_j \in \lbrace \mathcal{D} \cap \mathcal{U}_j \rbrace, \\ j \neq l, j \in \mathcal{L}}} x_{u_j}^{(n)} P_{j, u_j}^{(n)} g_{j, eNB}^{(n)} , & \hspace{-2em} $u_l \in \lbrace \mathcal{C} \cap \mathcal{U}_l \rbrace$ \nonumber \\ 
\displaystyle \sum_{\substack{\forall u_j \in  \mathcal{U}_j, \\ j \neq l, j \in \mathcal{L}}} x_{u_j}^{(n)}  P_{j, u_j}^{(n)} g_{j, u_l}^{(n)}, & \hspace{-3em} $ ~~ u_l \in \lbrace \mathcal{D} \cap \mathcal{U}_l \rbrace$. \nonumber   
\end{numcases} 

With the constraint in (\ref{eq:con-bin}), each RB is assigned to only one UE. With the constraints in (\ref{eq:con-pow-ue}) and (\ref{eq:con-pow-rel}), the transmit power is limited by the maximum power budget. The constraints in (\ref{eq:con-intf-1}) and (\ref{eq:con-intf-2}) limit the amount of interference introduced to the other relays and receiving D2D UEs in the first and second hop, respectively, to be less than some threshold. The constraint in (\ref{eq:con-QoS-cue}) ensures the minimum data rate requirements for the CUE and D2D UEs. The constraint in (\ref{eq:con-pow-0}) is the non-negativity condition for transmit power.

Similar to \cite{ref_user}, we apply the concept of reference node. As an example, in the first hop, each UE associated with relay node $l$ chooses from among the neighbouring relays having the highest channel gain according to following equation:
\begin{equation}
g_{{u_l^*}, l, 1}^{(n)} = \underset{j}{\operatorname{max}} ~ g_{u_l , j}^{(n)},~~ u_l \in \mathcal{U}_l, j \neq l, j \in \mathcal{L} \label{eq:ref_user1}
\end{equation}
and allocates the power level considering the interference threshold. Similarly, in the second hop, for each relay $l$, the transmit power   will be adjusted accordingly considering interference introduced to receiving D2D UEs (associated with neighboring relays) according to 
\begin{equation}
g_{l, {u_l^*}, 2}^{(n)} = \underset{u_j}{\operatorname{max}} ~ g_{l , u_j}^{(n)}, ~~ j \neq l, j \in \mathcal{L},  u_j \in \lbrace \mathcal{D} \cap \mathcal{U}_j \rbrace. \label{eq:ref_user2}
\end{equation}

From (\ref{eqn:e2e_rate}), the maximum rate for the UE $u_l$ over RB $n$ is achieved when $P_{u_l, l}^{(n)} \gamma_{u_l, l, 1}^{(n)} = P_{l, u_l}^{(n)} \gamma_{l, u_l, 2}^{(n)}$. Therefore, the power allocated to relay node for the UE $u_l$ can be expressed as a function of power at UE as $P_{l, u_l}^{(n)}  = \frac{\gamma_{u_l, l, 1}^{(n)}}{\gamma_{l, u_l, 2}^{(n)}}P_{u_l, l}^{(n)}$ and the rate of $u_l$ over RB $n$ is given by 
\begin{equation}
R_{u_l}^{(n)} = \frac{1}{2} B_{RB}  \log_2 \left( 1 + P_{u_l, l}^{(n)} \gamma_{u_l, l, 1}^{(n)} \right).
\end{equation}  
Hence the problem $\mathbf{P1}$ can be written as
\begin{subequations}
\setlength{\arraycolsep}{0.0em}
\begin{eqnarray}
 \mathbf{(P2)} 
\underset{x_{u_l}^{(n)}, P_{u_l, l}^{(n)}}{\operatorname{max}} ~ \sum_{u_l \in \mathcal{U}_l } \sum_{n =1}^N   \tfrac{1}{2}  x_{u_l}^{(n)}   B_{RB} \log_2  &&  \left(  1 +   P_{u_l, l}^{(n)} \gamma_{u_l, l, 1}^{(n)}   \right) \hspace{0.1em}   \nonumber \\
\text{subject to} \quad \sum_{u_l \in \mathcal{U}_l} x_{u_l}^{(n)} ~ &&{\leq} ~ 1, \quad ~~\forall n \hspace{-5em} \label{eq:con-bin-relx} \\
 \sum_{n =1}^N x_{u_l}^{(n)}  P_{u_l, l}^{(n)} ~ &&{\leq} ~ P_{u_l}^{max}, \forall u_l  \label{eq:con-pow-ue-relx} \\
 \sum_{u_l \in \mathcal{U}_l } \sum_{n =1}^N x_{u_l}^{(n)}  \tfrac{\gamma_{u_l, l, 1}^{(n)}}{\gamma_{l, u_l, 2}^{(n)}} P_{u_l, l}^{(n)} ~ &&{\leq} ~ P_l^{max} \label{eq:con-pow-rel-relx} \\
 \sum_{u_l \in \mathcal{U}_l } x_{u_l}^{(n)}  P_{u_l, l}^{(n)} g_{{u_l^*}, l, 1}^{(n)} ~ &&{\leq} ~ I_{th, 1}^{(n)}, ~\forall n \label{eq:con-intf-1-relx}\\
 \sum_{u_l \in \mathcal{U}_l } x_{u_l}^{(n)}  \tfrac{\gamma_{u_l, l, 1}^{(n)}}{\gamma_{l, u_l, 2}^{(n)}} P_{u_l, l}^{(n)} g_{l, {u_l^*}, 2}^{(n)} ~ &&{\leq} ~ I_{th, 2}^{(n)}, ~\forall n \label{eq:con-intf-2-relx} \\
\quad \sum_{n=1}^N \tfrac{1}{2}  x_{u_l}^{(n)} B_{RB} \log_2 \left(  1 + P_{u_l, l}^{(n)} \gamma_{u_l, l, 1}^{(n)} \right)  ~ &&{\geq} ~ Q_{u_l},   ~~\forall u_l   \label{eq:con-QoS-cue-relx}\\
\quad P_{u_l, l}^{(n)}  ~ &&{\geq} ~ 0, ~~ \forall n, u_l. \label{eq:con-pow-0-relx} 
\end{eqnarray}
\end{subequations}

\begin{remark}
\label{rem:MINLP}
The RAP formulation is a mixed-integer non-linear program (MINLP).  MINLP problems have the difficulties of both of their sub-classes, i.e., the combinatorial nature of mixed integer programs (MIPs) and the difficulty in solving nonlinear programs (NLPs). Since MIPs and NLPs are $\mathcal{NP}$-complete, the RAP $\mathbf{P2}$ is strongly  $\mathcal{NP}$-hard.
\end{remark} 

In order to obtain a tractable solution for the RAP formulation, in the following, we utilize the MP strategy.

\section{Message Passing Approach to Solve the Resource Allocation Problem} \label{sec:rap_mp}

\subsection{MP Strategy for the Max-sum Problem} \label{sec:MP_intro}

Given the RAP formulation $\mathbf{P2}$, we focus on the max-sum variant \cite{max-sum} of MP paradigm. Let us consider a generic function $f(y_1, y_2, \ldots,  y_J) : \mathfrak{D}_{\mathbf{y}} \rightarrow \mathbb{R}$ where each variable $y_j$ corresponds to a finite alphabet $\mathfrak{a}$, i.e., $\mathfrak{D}_{\mathbf{y}} = \mathfrak{a}^J$. We concentrate on maximizing the function $f(\cdot)$, i.e.,
\begin{equation} \label{eq:max_1}
\tilde{Z} = \underset{\mathbf{y}}{\operatorname{max}} ~ f(\mathbf{y}).
\end{equation} 
That is, $\tilde{Z}$ represents the maximization over all possible combinations of the vector $\mathbf{y} \in \mathfrak{a}^J$ where $\mathbf{y} = [y_1, y_2, \ldots,  y_J]^\mathsf{T}$. The \textit{marginal} of $\tilde{Z}$ with respect to variable $y_j$ is given by
\begin{equation} \label{eq:marginal_1}
\phi_{j}(y_j) = \underset{\sim (y_j)}{\operatorname{max}} ~ f(\mathbf{y})
\end{equation}
where $\underset{\sim (\alpha)}{\operatorname{max}} ~ f(\cdot)$ denotes the maximization over all variables in $f(\cdot)$ except variable $\alpha$. Let us decompose $f(\mathbf{y})$ into the summation of $K$ functions
$f_k(\cdot) : \mathfrak{D}_{\widehat{y}_k} \rightarrow \mathbb{R}, k \in \lbrace 1, 2, \ldots, K \rbrace $, i.e., $f(\mathbf{y}) = \displaystyle \sum_{k=1}^{K} f_k(\widehat{y}_k)$, where $\widehat{y}_k$ is a subset of elements of $\mathbf{y}$ and $\mathfrak{D}_{\widehat{y}_k} \subset \mathfrak{D}_{\mathbf{y}}$. Besides, let $\mathbf{f}\boldsymbol{(\cdot)} = \left[f_1(\cdot), f_2(\cdot), \ldots, f_K(\cdot) \right]^\mathsf{T}$ denote the vector of $K$ functions and $\mathfrak{f}_j$ represent the subset of functions in $\mathbf{f}\boldsymbol{(\cdot)}$ where the variable $y_j$ appears.  Hence, (\ref{eq:marginal_1}) can be rewritten as
\begin{equation}
\phi_{j}(y_j) = \underset{\sim (y_j)}{\operatorname{max}} ~ \sum_{k=1}^{K} f_k(\widehat{y}_k).
\end{equation}

Utilizing any MP algorithm, the computation of marginals involves passing messages between nodes represented by a specific graphical model. Among different graphical models, in this work, we consider factor graph \cite{bp_factor_graph} to capture the structure of generic function $f(\cdot)$. The factor graph consists of two different types of nodes, namely, function (or factor) nodes and variable nodes. A function node is connected with a variable node if and only if the variable appears in the corresponding function. Consequently, a factor graph contains two types of messages, i.e., message from factor nodes to variable nodes and vice-versa. According to the max-sum MP strategy, the message passed by any variable node $y_j, ~ j \in \lbrace 1, 2, \ldots, J \rbrace$, to any generic function node $f_k(\cdot), ~ k \in \lbrace 1, 2, \ldots, K \rbrace$, is given as
\begin{equation} \label{eq:msg_var_2_fact}
\delta_{y_j \rightarrow f_k(\cdot)}(y_j) = \sum_{\substack{i \in \mathfrak{f}_j, \\  i \neq k}} \delta_{f_i(\cdot) \rightarrow y_j} (y_j). 
\end{equation}
Likewise, the message from factor node $f_k(\cdot)$ to variable node $y_j$ is given as follows:
\begin{equation} \label{eq:msg_fact_2_var}
\delta_{f_k(\cdot) \rightarrow y_j } (y_j) = \underset{\sim (y_j)}{\operatorname{max}} \left( f_k(y_1, \ldots, y_J) + \hspace{-0.3em} \sum_{\substack{i \in \widehat{y}_k, \\ i \neq j}} \delta_{y_i \rightarrow f_k(\cdot)} (y_i) \right).
\end{equation}

When the factor graph is cycle free, it is represented as a tree (i.e., there is a unique path connecting any two nodes); hence, all the variable nodes can compute the marginals as
\begin{equation}
\phi_{j}(y_j) = \sum_{k = 1}^{K} \delta_{f_k(\cdot) \rightarrow y_j} (y_j).
\end{equation}
By invoking the general distributive law (i.e., $\max \sum = \sum \max$) \cite{mp_distributive}, the maximization in (\ref{eq:max_1}) can be computed as \begin{equation}
\tilde{Z} = \displaystyle \sum_{j=1}^{J} \underset{y_j}{\operatorname{max}} ~ \phi_{j}(y_j).
\end{equation}

\subsection{Utility Functions}

\begin{figure}[!t]
\centering
\includegraphics[scale=0.70]{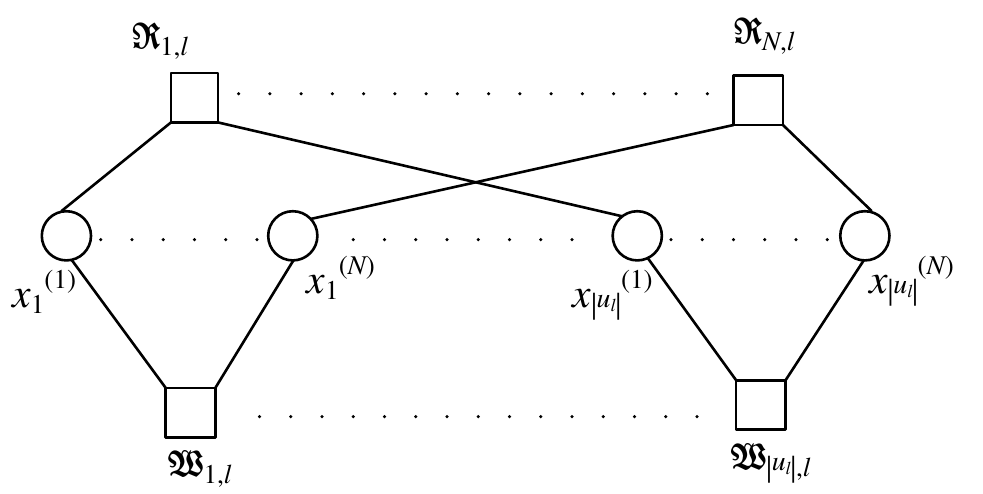}
\caption{An arbitrary factor graph representing MP formulation of the RAP. For ease of representation, the variables are denoted by circular nodes whereas the functions are denoted by square nodes. A variable node $x_{u_l}^{(n)}$ is connected to the function nodes $\mathfrak{R}_{n,l}(\cdot)$ and $\mathfrak{W}_{u_l, l}(\cdot)$ if and only if the variable appears in the corresponding function.} 
\label{fig:factorgraph}
\end{figure}

In the following, we develop a joint RB and power allocation mechanism that leverages the dynamics of MP strategy. Compared to centralized optimization solutions, MP allows to distribute the computational burden of achieving a feasible resource allocation by exchanging information among UEs and the corresponding relay.

In order to solve RAP $\mathbf{P2}$ using the MP scheme, we reformulate it as a utility maximization (i.e., cost minimization) problem and define the utility functions as in (\ref{eq:cost_function1}) and (\ref{eq:cost_function2}) where unfulfilled constraints result in infinite cost. Per RB constraints [i.e., (\ref{eq:con-bin-relx}), (\ref{eq:con-pow-rel-relx}), (\ref{eq:con-intf-1-relx}), (\ref{eq:con-intf-2-relx})] are incorporated in the utility function $\mathfrak{R}_{n,l}(\cdot)$ as follows:
\begin{equation}
\label{eq:cost_function1}
  \hspace{-0.15em}\mathfrak{R}_{n,l}(\cdot)  = \left\{\begin{alignedat}{2}
    & 0, & \text{if} ~~ \displaystyle \sum_{u_l \in \mathcal{U}_l} x_{u_l}^{(n)} & \leq 1 \\
    & & \displaystyle \sum_{u_l \in \mathcal{U}_l }  x_{u_l}^{(n)} \tfrac{\gamma_{u_l, l, 1}^{(n)}}{\gamma_{l, u_l, 2}^{(n)}} P_{u_l, l}^{(n)} & \leq  {P_l^{(n)}}^{max} \\
    & & \displaystyle  \sum_{u_l \in \mathcal{U}_l } x_{u_l}^{(n)} P_{u_l, l}^{(n)} g_{{u_l^*}, l, 1}^{(n)}  &\leq I_{th, 1}^{(n)} \\
    &  & \displaystyle  \quad \sum_{u_l \in \mathcal{U}_l } x_{u_l}^{(n)}  \tfrac{\gamma_{u_l, l, 1}^{(n)}}{\gamma_{l, u_l, 2}^{(n)}} P_{u_l, l}^{(n)}  g_{l, {u_l^*}, 2}^{(n)} & \leq I_{th, 2}^{(n)} \\
    & - \infty, & \text{otherwise} \hspace{2em}
  \end{alignedat}\right.
\end{equation}
where $ {P_l^{(n)}}^{max} = \frac{ P_l^{max} }{N}$. On the other hand, per UE constraints are incorporated in the utility function $\mathfrak{W}_{u_l, l}(\cdot)$ which is the achievable rate of each UE only if the constraints in (\ref{eq:con-pow-ue-relx}) and (\ref{eq:con-QoS-cue-relx}) are satisfied, i.e., 
\begin{equation}
\label{eq:cost_function2}
\mathfrak{W}_{u_l, l}(\cdot)  = \left\{\begin{alignedat}{2}
    &   \sum_{n =1}^N   x_{u_l}^{(n)} R_{u_l}^{(n)}, & \text{if} ~~ \displaystyle \sum_{n =1}^N x_{u_l}^{(n)} P_{u_l, l}^{(n)} & \leq  P_{u_l}^{max} \\
    & & \sum_{n =1}^N   x_{u_l}^{(n)} R_{u_l}^{(n)}  & \geq  Q_{u_l}\\
    & - \infty, & \text{otherwise.} \hspace{5em}
  \end{alignedat}\right.
\end{equation}

\subsection{MP Formulation for the Resource Allocation Problem}

Using the utility functions above, the RAP for each relay $l$ can be rewritten as
\begin{equation} \label{eq:margin_RB}
\mathbf{x}_{\boldsymbol l}^* = \underset{\mathbf{x}}{\operatorname{max}} \left( \sum_{n=1}^{N} \mathfrak{R}_{n,l}(\cdot) + \sum_{u_l \in \mathcal{U}_l} \mathfrak{W}_{u_l, l}(\cdot) \right).
\end{equation}

By exploiting the concept  described in Section \ref{sec:MP_intro}, let us associate (\ref{eq:margin_RB}) with a factor graph as shown in Fig. \ref{fig:factorgraph}. Following an MP strategy, the variable and function nodes exchange messages along their connecting edges until the values of $x_{u_l}^{(n)}$  are determined for $\forall u_l, n$. Let $\phi_{u_l}^{(n)}$ be the marginalization of (\ref{eq:margin_RB}) with respect to $x_{u_l}^{(n)}$ and given as
\begin{equation}
\phi_{u_l}^{(n)}\left( x_{u_l}^{(n)} \right) = \underset{\sim \left(x_{u_l}^{(n)} \right)}{\operatorname{max}} \left( \sum_{n=1}^{N} \mathfrak{R}_{n,l}(\cdot) + \sum_{u_l \in \mathcal{U}_l} \mathfrak{W}_{u_l, l}(\cdot) \right).
\end{equation}

Let $\delta_{\mathfrak{R}_{n,l}(\cdot) \rightarrow x_{u_l}^{(n)}} \left( x_{u_l}^{(n)} \right)$  and $\delta_{x_{u_l}^{(n)} \rightarrow \mathfrak{R}_{n,l}(\cdot)} \left( x_{u_l}^{(n)} \right) $ denote
 the message exchanged between  function nodes $\mathfrak{R}_{n,l}(\cdot)$ and the connected variable nodes for $\forall u_l, n$. Similarly, $ \delta_{\mathfrak{W}_{u_l, l}(\cdot)  \rightarrow x_{u_l}^{(n)}} \left( x_{u_l}^{(n)} \right) $
and $\delta_{x_{u_l}^{(n)} \rightarrow \mathfrak{W}_{u_l, l}(\cdot)} \left( x_{u_l}^{(n)} \right)$ denote the  exchanged messages between  function nodes $ \mathfrak{W}_{u_l, l}(\cdot) $ and variable nodes for $\forall u_l, n$. Let us consider a generic RB $n$ in the factor graph. The square node in Fig. \ref{fig:factorgraph} corresponding  to $\mathfrak{R}_{n,l}(\cdot)$ which is connected to all variable nodes $x_{u_l}^{(n)}$ for $\forall u_l \in \mathcal{U}_l$. Hence from (\ref{eq:msg_fact_2_var}), the message to be delivered to the particular variable node $x_{u_l}^{(n)}$ is obtained as follows:
\begin{align} \label{eq:message1}
\delta_{\mathfrak{R}_{n,l}(\cdot) \rightarrow x_{u_l}^{(n)}} \left( x_{u_l}^{(n)} \right) =  \max \sum_{j \in \mathcal{U}_l, j \neq u_l } & \delta_{x_{j}^{(n)} \rightarrow \mathfrak{R}_{n,l}(\cdot)  } \left( x_{j}^{(n)} \right) \nonumber \\
\text{subject to } 
\displaystyle \sum_{u_l \in \mathcal{U}_l} x_{u_l}^{(n)} & \leq 1 \nonumber  \\
\displaystyle \sum_{u_l \in \mathcal{U}_l }  x_{u_l}^{(n)} \tfrac{\gamma_{u_l, l, 1}^{(n)}}{\gamma_{l, u_l, 2}^{(n)}} P_{u_l, l}^{(n)} & \leq  {P_l^{(n)}}^{max} \nonumber  \\
\displaystyle  \sum_{u_l \in \mathcal{U}_l } x_{u_l}^{(n)} P_{u_l, l}^{(n)} g_{{u_l^*}, l, 1}^{(n)}  &\leq I_{th, 1}^{(n)} \nonumber  \\
\displaystyle  \quad \sum_{u_l \in \mathcal{U}_l } x_{u_l}^{(n)}  \tfrac{\gamma_{u_l, l, 1}^{(n)}}{\gamma_{l, u_l, 2}^{(n)}} P_{u_l, l}^{(n)} g_{l, {u_l^*}, 2}^{(n)} & \leq I_{th, 2}^{(n)}.
\end{align}

Let us consider a generic user $u_l$. As illustrated in Fig. \ref{fig:factorgraph}, the square nodes corresponding to function $\mathfrak{W}_{u_l, l}(\cdot)$ in factor graph are connected to all variable nodes $x_{u_l}^{(n)}$ for $\forall n \in \mathcal{N}$. Using (\ref{eq:msg_fact_2_var}) and (\ref{eq:cost_function2}),  the message from  function node $\mathfrak{W}_{u_l, l}(\cdot)$ to any variable node $x_{u_l}^{(n)}$ is given by (\ref{eq:message2}).

\begin{figure*}[!t]
\normalsize

\begin{align} \label{eq:message2}
\delta_{\mathfrak{W}_{u_l, l}(\cdot)  \rightarrow x_{u_l}^{(n)}} \left( x_{u_l}^{(n)} \right) =  x_{u_l}^{(n)} R_{u_l}^{(n)}   &+    
\max \left( \sum_{\substack{j =1, \\ j \neq n}}^N   x_{u_l}^{(j)} R_{u_l}^{(j)}  + \delta_{x_{u_l}^{(j)} \rightarrow \mathfrak{W}_{u_l, l}(\cdot)   } \left( x_{u_l}^{(j)} \right) \right) \nonumber \\
\text{subject to } \displaystyle \sum_{n =1}^N x_{u_l}^{(n)} P_{u_l, l}^{(n)} & \leq  P_{u_l}^{max}, \quad \quad
    \sum_{n =1}^N    x_{u_l}^{(n)} R_{u_l}^{(n)}   \geq  Q_{u_l}.
\end{align}

\hrulefill
\vspace*{4pt}
\end{figure*}

From (\ref{eq:message1}) and (\ref{eq:message2}), the marginal  $\phi_{u_l}^{(n)}\left( x_{u_l}^{(n)} \right)$ can be obtained as
\begin{equation}
\phi_{u_l}^{(n)}\left( x_{u_l}^{(n)} \right) = \delta_{\mathfrak{R}_{n,l}(\cdot) \rightarrow x_{u_l}^{(n)}} \left( x_{u_l}^{(n)} \right) + \delta_{\mathfrak{W}_{u_l, l}(\cdot)  \rightarrow x_{u_l}^{(n)}} \left( x_{u_l}^{(n)} \right).
\end{equation}
Consequently, the RB allocation indicator for UE $u_l$ over RB $n$ is given by
\begin{equation} \label{eq:x_opt}
{x_{u_l}^{(n)}}^* =  \underset{ x_{u_l}^{(n)} }{\operatorname{argmax}}  \left[ \phi_{u_l}^{(n)}\left( x_{u_l}^{(n)} \right) \right]. 
\end{equation}

From (\ref{eq:message1}) and (\ref{eq:message2}),  it can be noted that both the messages, i.e., $\delta_{\mathfrak{R}_{n,l}(\cdot) \rightarrow x_{u_l}^{(n)}} \left( x_{u_l}^{(n)} \right)$ and $\delta_{\mathfrak{W}_{u_l, l}(\cdot)  \rightarrow x_{u_l}^{(n)}} \left( x_{u_l}^{(n)} \right)$ solve a local optimization problem with respect to the allocation variable $ x_{u_l}^{(n)}$. It is worth noting that, in our system model,  each  function node $\mathfrak{W}_{u_l, l}(\cdot)$ and corresponding variable nodes are located at the UE $u_l$, while all $\delta_{\mathfrak{R}_{n,l}}(\cdot) $ nodes are located at the relay. Hence, sending messages $\delta_{\mathfrak{R}_{n,l}}(\cdot) $ from variable nodes to  function nodes (and vice-versa) requires actual transmission on the radio channel. However, the message exchanges between variable nodes and  function nodes $\mathfrak{W}_{u_l, l}(\cdot)$ are performed locally at the UEs without actual transmission on the radio channel.

\subsection{An Effective Implementation of MP Strategy}

In a practical LTE-A system, since the exchange of messages actually involves effective transmissions over the channel, the MP scheme described in the preceding section might be limited by the signaling overhead  due to transfer of messages between relay and UEs. In the following, we observe that the amount of message signaling can be significantly reduced by some algebraic manipulations. Note that, the message $\delta_{\mathfrak{W}_{u_l, l}(\cdot)  \rightarrow x_{u_l}^{(n)}} \left( 1\right)$ carries information regarding the use of RB $n$ by UE $u_l$ with transmission power $P_{u_l,l}^{(n)}$, while $\delta_{\mathfrak{W}_{u_l, l}(\cdot)  \rightarrow x_{u_l}^{(n)}} \left( 0 \right)$ carries information regarding the lack of transmission on RB $n$ by UE $u_l$, i.e., $P_{u_l,l}^{(n)} = 0$. Hence, each UE eventually delivers a real-valued vector of two elements, i.e., $$\boldsymbol{ \Delta_{\mathfrak{W}_{u_l, l}(\cdot)  \rightarrow x_{u_l}^{(n)}} } = \left[ \delta_{\mathfrak{W}_{u_l, l}(\cdot)  \rightarrow x_{u_l}^{(n)}} \left( 1\right), \delta_{\mathfrak{W}_{u_l, l}(\cdot)  \rightarrow x_{u_l}^{(n)}} \left( 0\right)  \right]^\mathsf{T}.$$ Let $\kappa_{u_l}$ denote the required number of RB(s)\footnote{The calculation of $\kappa_{u_l}$ is given in \textbf{Appendix \ref{app:num_rb}}.} to satisfy the data rate requirement $Q_{u_l}$ for UE $u_l$. Therefore, the constraint in (\ref{eq:con-QoS-cue-relx}) can be rewritten as 
\begin{equation} \label{eq:new_qos}
\sum_{n =1}^N    x_{u_l}^{(n)}   \geq  \kappa_{u_l}, ~~ \forall u_l.     
\end{equation}
Now, replacing the constraint in (\ref{eq:message2}) with that in (\ref{eq:new_qos}) and subtracting the constant term $\displaystyle \sum_{\substack{j =1; \\ j \neq n}}^N \delta_{x_{u_l}^{(j)} \rightarrow \mathfrak{W}_{u_l, l}(\cdot)   } \left(0 \right) $ from both sides of (\ref{eq:message2}), we obtain (\ref{eq:message_modified}). Let us introduce the normalized messages $\tilde{\psi}_{u_l, l}^{(n)} = \delta_{x_{u_l}^{(n)} \rightarrow \mathfrak{W}_{u_l, l}(\cdot)   } \left( 1 \right) - \delta_{x_{u_l}^{(n)} \rightarrow \mathfrak{W}_{u_l, l}(\cdot)   } \left(0 \right) = \delta_{\mathfrak{R}_{n,l}(\cdot) \rightarrow x_{u_l}^{(n)}} \left( 1 \right) - \delta_{\mathfrak{R}_{n,l}(\cdot) \rightarrow x_{u_l}^{(n)}} \left( 0 \right)$. 
It can be observed that the terms within the summation in (\ref{eq:message_modified}) are either $0$ or $R_{u_l}^{(n)} + \tilde{\psi}_{u_l, l}^{(n)}$ depending on whether the RB allocation indicator variable $x_{u_l}^{(n)}$ is $0$ or $1$. 

\begin{figure*}[!t]
\normalsize

\begin{align} \label{eq:message_modified}
\delta_{\mathfrak{W}_{u_l, l}(\cdot)  \rightarrow x_{u_l}^{(n)}} \left( x_{u_l}^{(n)} \right) - \sum_{\substack{j =1; \\ j \neq n}}^N \delta_{x_{u_l}^{(j)} \rightarrow \mathfrak{W}_{u_l, l}(\cdot)   } \left(0 \right) =  x_{u_l}^{(n)} R_{u_l}^{(n)}  & +   
\max \left( \sum_{\substack{j =1, \\ j \neq n}}^N   x_{u_l}^{(j)} R_{u_l}^{(j)}  + \delta_{x_{u_l}^{(j)} \rightarrow \mathfrak{W}_{u_l, l}(\cdot)   } \left( x_{u_l}^{(j)} \right) - \delta_{x_{u_l}^{(j)} \rightarrow \mathfrak{W}_{u_l, l}(\cdot)   } \left(0 \right) \right) \nonumber \\
\text{subject to } \displaystyle \sum_{n =1}^N x_{u_l}^{(n)} P_{u_l, l}^{(n)} & \leq  P_{u_l}^{max}, \quad \quad
    \sum_{n =1}^N    x_{u_l}^{(n)}    \geq  \kappa_{u_l}.
\end{align}

\hrulefill
\vspace*{4pt}
\end{figure*}

Given the above, the maximization is straightforward. For instance, consider the vector  $$\boldsymbol {\chi_{u_l}} = \left[ R_{u_l}^{(1)} + \tilde{\psi}_{u_l, l}^{(1)}, \ldots, R_{u_l}^{(j)} + \tilde{\psi}_{u_l, l}^{(j)}, \ldots, R_{u_l}^{(N)} + \tilde{\psi}_{u_l, l}^{(N)} \right]^\mathsf{T}$$ and $\langle \upsilon_{u_l}^{(j)}\rangle_{z \setminus n}$ be the $z$-th sorted element of $\boldsymbol {\chi_{u_l}}$ without considering the term $R_{u_l}^{(j)} + \tilde{\psi}_{u_l, l}^{(j)}$ so that $$\langle \upsilon_{u_l}^{(j)}\rangle_{(z-1) \setminus n} \geq \langle \upsilon_{u_l}^{(j)}\rangle_{z \setminus n} \geq \langle \upsilon_{u_l}^{(j)}\rangle_{(z+1) \setminus n}$$ for $\forall j \in \mathcal{N}, j \neq n$. Hence, for $x_{u_l}^{(n)} =1$, the maximum rate will be achieved if \cite{mp-twireless}
\begin{align} \label{eq:msg_new_1}
\delta_{\mathfrak{W}_{u_l, l}(\cdot)  \rightarrow x_{u_l}^{(n)}} \left( 1 \right) -&  \sum_{\substack{j =1, \\ j \neq n}}^N \delta_{x_{u_l}^{(j)} \rightarrow \mathfrak{W}_{u_l, l}(\cdot)   } \left(0 \right)  \hspace{2em} \nonumber \\ =& ~~ R_{u_l}^{(n)} +  \sum_{z=1}^{\kappa_{u_l} - 1} \langle \upsilon_{u_l}^{(j)}\rangle_{z \setminus n}.
\end{align}
Similarly, for $x_{u_l}^{(n)} =0$, the maximum is given by \cite{mp-twireless}
\begin{equation} \label{eq:msg_new_2}
\delta_{\mathfrak{W}_{u_l, l}(\cdot)  \rightarrow x_{u_l}^{(n)}} \left( 0 \right) - \sum_{\substack{j =1; \\ j \neq n}}^N \delta_{x_{u_l}^{(j)} \rightarrow \mathfrak{W}_{u_l, l}(\cdot)   } \left(0 \right) = \sum_{z=1}^{\kappa_{u_l}} \langle \upsilon_{u_l}^{(j)}\rangle_{z \setminus n}.
\end{equation}
Since by definition $$\psi_{u_l, l}^{(n)} = \delta_{\mathfrak{W}_{u_l, l}(\cdot) \rightarrow x_{u_l}^{(n)}     } \left( 1 \right) - \delta_{\mathfrak{W}_{u_l, l}(\cdot) \rightarrow  x_{u_l}^{(n)}    } \left(0 \right),$$ from (\ref{eq:msg_new_1}) and (\ref{eq:msg_new_2}), the normalized messages can be derived as follows:
\begin{align} \label{eq:message_ue2rb}
\psi_{u_l, l}^{(n)} &= R_{u_l}^{(n)} - \langle \upsilon_{u_l}^{(j)}\rangle_{\kappa_{u_l} \setminus n} \nonumber \\
&= R_{u_l}^{(n)} - \langle R_{u_l}^{(j)} + \tilde{\psi}_{u_l, l}^{(j)} \rangle_{\kappa_{u_l} \setminus n}
\end{align}
where $j \in \mathcal{N}$ and $j \neq n$.  Note that the messages sent from UE $u_l$ to RB $n$ in factor graph is a scalar quantity. Similarly, the normalized messages from RB $n$ to UE $u_l$, i.e., $\tilde{\psi}_{u_l, l}^{(n)} = \delta_{\mathfrak{R}_{n,l}(\cdot) \rightarrow x_{u_l}^{(n)}} \left( 1 \right) - \delta_{\mathfrak{R}_{n,l}(\cdot) \rightarrow x_{u_l}^{(n)}} \left( 0 \right)$ becomes \cite{mp-twireless}
\begin{equation} \label{eq:message_rb2ue}
\tilde{\psi}_{u_l, l}^{(n)} = - ~\underset{\substack{i \in \mathcal{U}_l, \\  i \neq u_l}}{\operatorname \max} ~~ \psi_{i, l}^{(n)}.
\end{equation} 

Note that, for any arbitrary graph, the allocation variables may keep oscillating and might not converge to any fixed point. In the context of loopy graphical models, by introducing a suitable weight, the messages in (\ref{eq:message_ue2rb}) and (\ref{eq:message_rb2ue}) perturb to a fixed point. Accordingly, (\ref{eq:message_ue2rb}) and (\ref{eq:message_rb2ue}) can be rewritten as \cite{min-sum-mp}
\begin{subequations}
\begin{equation} \label{eq:message_ue2rb_weight}
  \psi_{u_l, l}^{(n)} = R_{u_l}^{(n)} - \omega \left\langle R_{u_l}^{(j)} + \psi_{u_l, l}^{(j)} \right\rangle_{\kappa_{u_l} \setminus n} + (1 - \omega) \left( R_{u_l}^{(n)} + \tilde{\psi}_{u_l, l}^{(n)} \right) 
\end{equation}    
\begin{equation} \label{eq:message_rb2ue_weight}
\tilde{\psi}_{u_l, l}^{(n)} = - \omega ~\underset{\substack{i \in \mathcal{U}_l, \\  i \neq u_l}}{\operatorname \max} ~~ \psi_{i, l}^{(n)} - (1 - \omega) \psi_{u_l, l}^{(n)}.
\end{equation}
\end{subequations}
Note that, when $\omega =1$, (\ref{eq:message_ue2rb_weight}) and (\ref{eq:message_rb2ue_weight}) reduce to the original formulation, i.e., (\ref{eq:message_ue2rb}) and (\ref{eq:message_rb2ue}), respectively. Thus the solution ${x_{u_l}^{(n)}}^*$ can be easily obtained by calculating the node marginals for each UE-RB pair, i.e., for all $u_l \in \mathcal{U}_l, n \in \mathcal{N}$ pair as follows:
\begin{equation}
\tau_{u_l, l}^{(n)} =  \psi_{u_l, l}^{(n)} + \tilde{\psi}_{u_l, l}^{(n)}.
\end{equation}
Hence, from (\ref{eq:x_opt}), the optimal RB allocation can be computed as
\begin{equation} \label{eq:rb_alloc_mp}
{x_{u_l}^{(n)}}^*  = \begin{cases}
 0, \quad  \text{if $\tau_{u_l, l}^{(n)} < 0$} \\
 1, \quad \text{otherwise.}
\end{cases}
\end{equation}

\section{Distributed Solution for the Resource Allocation Problem} \label{sec:distributed_sol}

\subsection{Algorithm Development}

Once the optimal RB allocation is obtained, the transmission power of the UEs on assigned RB(s) is obtained as follows. We couple the classical generalized distributed constrained power control scheme (GDCPC) \cite{power_gdpc} with an autonomous power control method \cite{auto_pow_cr} which considers the data rate requirements of UEs while protecting other receiving nodes from interference. More specifically, at each iteration, the transmission power is updated using  (\ref{eq:pow_alloc_mp}) where $${P_{u_l}^{(n)}}^{max} = \tfrac{ P_{u_l}^{max} } { \displaystyle \sum_{n=1}^{N} x_{u_l}^{(n)} }$$ and $ \hat{P}_{u_l,l}^{(n)}$ is obtained as 
\begin{align} \label{eq:power_p_tilde}
 \hat{P}_{u_l,l}^{(n)} = \min \left( \tilde{P}_{u_l,l}^{(n)}, ~ \min \left( {P_{u_l}^{(n)}}^{max}, \varpi_{u_l,l}^{(n)} \right) \right).
\end{align}

\begin{figure*}[!t]
\normalsize

\begin{align} \label{eq:pow_alloc_mp}
P_{u_l,l}^{(n)}(t+1) =   \begin{cases}
 \tfrac{2^{Q_{u_l}  } -1 }{2^{R_{u_l}(t) } - 1} P_{u_l,l}^{(n)}(t), \quad  \text{if $\tfrac{2^{Q_{u_l}  } -1 }{2^{R_{u_l}(t) } - 1} P_{u_l,l}^{(n)}(t) \leq {P_{u_l}^{(n)}}^{max}$} \vspace{0.5em}\\ 
 \hat{P}_{u_l,l}^{(n)}, \hspace{5.8em} \text{otherwise} 
\end{cases}
\end{align}

\hrulefill
\vspace*{4pt}
\end{figure*}

In (\ref{eq:power_p_tilde}), $\tilde{P}_{u_l,l}^{(n)}$ is chosen arbitrarily within the range of $0 \leq \tilde{P}_{u_l,l}^{(n)} \leq {P_{u_l}^{(n)}}^{max}$ and $\varpi_{u_l,l}^{(n)} $ is given by
\begin{equation}
\varpi_{u_l,l}^{(n)} =  \min \left( \tfrac{  I_{th, 1}^{(n)} } { g_{{u_l^*}, l, 1}^{(n)}  },  \tfrac{\gamma_{l, u_l, 2}^{(n)}}{\gamma_{u_l, l, 1}^{(n)}} \cdot \tfrac{ I_{th, 2}^{(n)} }{ g_{l, {u_l^*}, 2}^{(n)} }    \right).
\end{equation}

 Each relay independently performs the resource allocation and allocates
resources to the associated UEs. For completeness, the distributed joint RB and power allocation algorithm is summarized in \textbf{Algorithm \ref{alg:mp_rec_alloc}}. Since the L3 relays can perform the same operation as an eNB, these relays can communicate using the X2 interface
\cite{lte_arch} defined in the LTE-A specification. Therefore, the relays can obtain the channel state information through inter-relay message passing without increasing the overhead of signalling at the eNB.

\begin{algorithm*}
\caption{Allocation of RB and transmission power using message passing}
\label{alg:mp_rec_alloc}
\begin{algorithmic}[1]

\STATE Estimate channel quality indicator (CQI) matrices from previous time slot.
\STATE Initialize $t:= 0, P_{u_l,l}^{(n)}(0) := \frac{P_{u_l}^{max}}{N} , \psi_{u_l, l}^{(n)}(0) :=0, \tilde{\psi}_{u_l, l}^{(n)}(0) :=0$ for $\forall u_l \in \mathcal{U}_l, n \in \mathcal{N}$.

\REPEAT

\STATE Each UE $u_l$ sends messages $\psi_{u_l, l}^{(n)}(t+1) = R_{u_l}^{(n)}(t) - \omega \left\langle R_{u_l}^{(j)}(t) + \psi_{u_l, l}^{(j)}(t) \right\rangle_{\kappa_{u_l} \setminus n} + (1 - \omega) \left( R_{u_l}^{(n)}(t) + \tilde{\psi}_{u_l, l}^{(n)}(t) \right)$ to the relay $l \in \mathcal{L}$ for each RB $n \in \mathcal{N}$.

\STATE The relay $l \in \mathcal{L}$ sends messages $\tilde{\psi}_{u_l, l}^{(n)}(t+1) = - \omega ~\underset{\substack{i \in \mathcal{U}_l, \\  i \neq u_l}}{\operatorname \max} ~~ \psi_{i, l}^{(n)}(t) - (1 - \omega) \psi_{u_l, l}^{(n)}(t)$ to each associated UE $u_l \in \mathcal{U}_l$ for $\forall n \in \mathcal{N}$.

\STATE Each UE $u_l$ computes the marginals as $\tau_{u_l, l}^{(n)} (t+1) =  \psi_{u_l, l}^{(n)} (t) + \tilde{\psi}_{u_l, l}^{(n)} (t)$ for $\forall n \in \mathcal{N}$ and reports to the corresponding relay. 

\STATE Each relay $l$ calculates the RB and power allocation vector for each UE according to (\ref{eq:rb_alloc_mp}) and (\ref{eq:pow_alloc_mp}), respectively.

\STATE Calculate the aggregated achievable network rate as $R_l(t+1) := \displaystyle \sum_{u_l \in \mathcal{U}_l} R_{u_l}(t+1)$.   

\STATE Update $t:= t + 1$.

\UNTIL $t = T_{max}$ or the convergence criterion met (i.e., $ \mathbf{abs} \lbrace R_l(t+1) - R_l(t) \rbrace < \varepsilon$, where $\varepsilon$ is the  tolerance for convergence).

\STATE  Allocate resources (i.e., RB and transmit power) to the associated UEs for each relay. 

\end{algorithmic}
\end{algorithm*}

\begin{remark}
\label{rem:algo_bound}
Since ${\mathbf{x}_{\boldsymbol l}}^*$ satisfies the binary constraints, and the optimal allocation  $\left({\mathbf{x}_{\boldsymbol l}}^*, {\mathbf{P}_{\boldsymbol l}}^*\right)$ satisfies all the constraints in $\mathbf{P2}$, for a sufficient number of available RBs, the solution obtained by {\normalfont \textbf{Algorithm \ref{alg:mp_rec_alloc}} } gives a lower bound on the solution of the original RAP $\mathbf{P2}$.
\end{remark}

\subsection{Complexity Analysis}

If the algorithm requires $T$ iterations to converge, it is easy to verify that the time complexity at each relay $l \in \mathcal{L}$ is of $\mathcal{O}(T |\mathcal{U}_l| N)$. Similarly, considering a standard sorting algorithm (e.g., merge sort, heap sort) to generate the outputs $\langle \upsilon_{u_l}^{(j)}\rangle_{z \setminus n}$ for $\forall n$ with a worst-case complexity of $\mathcal{O}(N \log N)$, the overall time complexity at each UE is $\mathcal{O}\left( T N^2 \log N \right)$.

\subsection{Convergence of the Algorithm and Optimality of the Solution}

\begin{theorem}
\label{theorem:optimal-proof-mp}

If the algorithm converges to a fixed point message, this point follows the slackness condition of $\mathbf{P2}$, and hence it becomes the optimal solution for the original resource allocation problem.

\end{theorem}
\begin{IEEEproof}
See \textbf{Appendix \ref{app:optimal-mp}}.
\end{IEEEproof}

\begin{theorem}
\label{theorem:converge}

The message passing algorithm converges to a solution with zero duality gap as the number of resource blocks goes to infinity, i.e., dual problem of $\mathbf{P2}$ [e.g., $\mathscr{D}_l$, given by (\ref{eq:conv-mp2})] has the same optimal objective function value \cite{mp-dft}.
\end{theorem}
\begin{IEEEproof}
See \textbf{Appendix \ref{app:converge-mp}}.
\end{IEEEproof}

\subsection{End-to-End Delay for the Proposed Solution}

We measure the total end-to-end delay due to relaying for the proposed framework as follows \cite{delay_imt}:
\begin{equation} \label{eq:delay}
\mathfrak{D}_{\rm 2 hop} = \mathfrak{t}_{\rm schedule} + \mathfrak{t}_{\rm delivery}^{[1]} + \mathfrak{t}_{\rm decode}  + \mathfrak{t}_{\rm delivery}^{[2]}
\end{equation}
where  $\mathfrak{t}_{\rm schedule}$ is the time required to schedule the UEs and perform resource allocation, $\mathfrak{t}_{\rm decode}$ is the decoding time at relay nodes before data packets are forwarded in second hop, and $\mathfrak{t}_{\rm delivery}^{[j]} = \mathfrak{t}_{\rm transmit}^{[j]} + \mathfrak{t}_{\rm pd}^{[j]}$ is the sum of packet transmission time and propagation delay for hop $j \in \lbrace 1,2 \rbrace$. While calculating delay using  (\ref{eq:delay}), we assume that each scheduled UE is ready to transmit data and the waiting time before transmission is zero (i.e., there is no queuing delay).

\subsection{Implementation of Proposed Solution in a Practical LTE-A Scenario}

\begin{figure}[!t]
\centering
\includegraphics[width=1.7in]{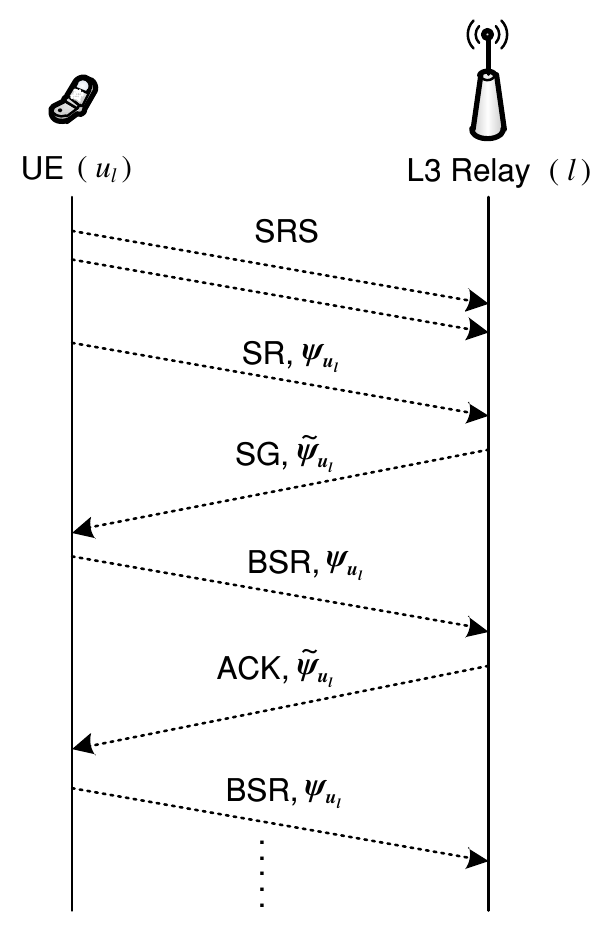}
\caption{Possible implementation of the MP scheme in an LTE-A system.} 
\label{fig:mp_signal}
\end{figure}

Let $\boldsymbol{\psi_{u_l}} = \left[ \psi_{u_l}^{(1)}, \psi_{u_l}^{(2)}, \ldots, \psi_{u_l}^{(N)}\right]^\mathsf{T}$ and $\boldsymbol{\tilde{\psi}_{u_l}} = \left[ \tilde{\psi}_{u_l}^{(1)}, \tilde{\psi}_{u_l}^{(2)}, \ldots, \tilde{\psi}_{u_l}^{(N)} \right]^\mathsf{T}$ denote the  message vectors for UE $u_l$. These messages can be mapped into standard LTE-A scheduling control messages as illustrated in Fig. \ref{fig:mp_signal}. In an LTE-A system, UEs periodically sense the physical uplink control channel (PUCCH) and transmit known sequences using sounding reference signal (SRS). After reception of scheduling request (SR) from UEs, an L3 relay performs scheduling and resource allocation. After scheduling, the L3 relay allocates RB(s) and informs to the UEs by sending scheduling grant (SG) through  physical downlink control channel (PDCCH). Once the allocation of RB(s) is received, the UEs periodically send the buffer status report (BSR) using PUCCH to the relay in order to update the resource requirement, and in response, the relay sends back an acknowledgment (ACK) in physical hybrid-ARQ indicator channel (PHICH). Considering the above scenario, our proposed message passing approach can be implemented by incorporating $\boldsymbol{\psi_{u_l}}$ messages in SR and BSR, and $\boldsymbol{\tilde{\psi}_{u_l}}$ messages in SG and ACK control signals, respectively.

\section{Performance Evaluation} \label{sec:performance}

\subsection{Simulation Setup}

\begin{figure}[!h t b]
\centering
\includegraphics[width=1.7in]{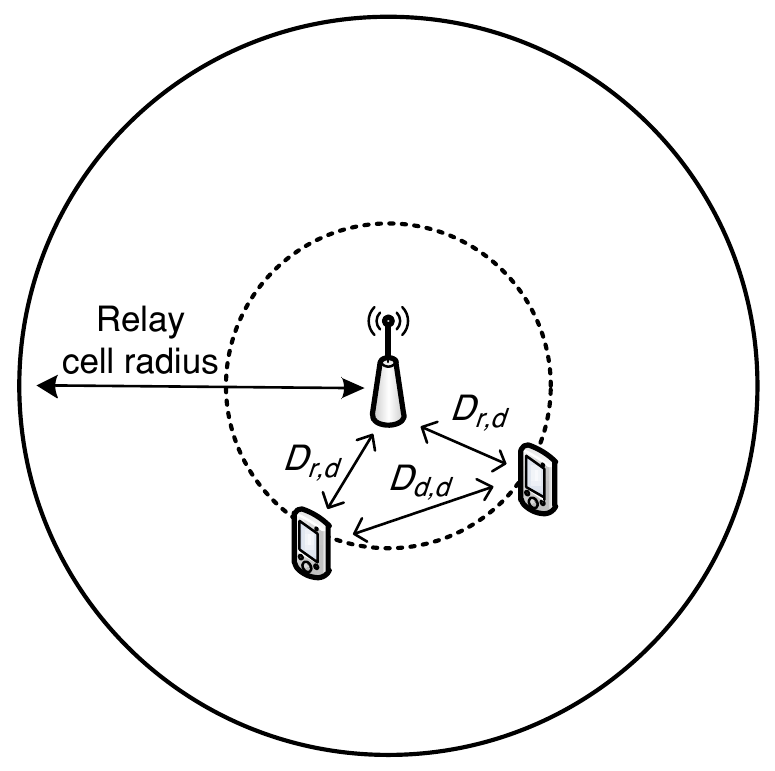}
\caption{D2D UEs are uniformly distributed within the radius $D_{r,d}$ while keeping distance $D_{d,d}$ between peers.} 
\label{fig:d2d_position}
\end{figure}

In order to evaluate the performance of the proposed resource allocation scheme, we develop an event-driven simulator. All the simulations are performed in MATLAB environment. The simulator focuses on capturing the medium access control (MAC) layer behavior of the LTE-A network. We simulate a single three-sectored cell in a rectangular area of $700~\text{m} \times 700~\text{m}$, where the eNB is located in the center of the cell and three relays are deployed in the network, i.e., one relay in each sector. The CUEs are uniformly distributed within the relay cell. The D2D transmitters and receivers are uniformly distributed within a radius $D_{r,d}$ while keeping a distance $D_{d,d}$ between peers as shown in Fig. \ref{fig:d2d_position}.  Both $D_{r,d}$ and $D_{d,d}$ are varied as  simulation parameters. We consider a snapshot model to obtain the network performance, where all the network parameters remain constant during a simulation run.

\begin{table}[!t]
\renewcommand{\arraystretch}{1.3}
\caption{Simulation Parameters}
\label{tab:sim_param}
\centering
\begin{tabular}{l|l}
\hline
\bfseries Parameter & \bfseries Values\\
\hline\hline
Cell layout & Hexagonal grid, $3$-sector sites \\
Carrier frequency & $2.35$ GHz \\
System bandwidth & $2.5$ MHz \\
Total number of available RBs & $13$ \\
MAC frame duration & $10$ msec \\
Scheduling time & $0.10$ msec \\
Packet size & $1500$ bytes \\
Relay cell radius & $200$ meter\\
Distance between eNB and relays & $125$ meter\\
Minimum distance between UE and relay & $10$ meter\\
Total power available at each relay & $30$ dBm \\
Total power available at UE & $23$ dBm \\
Rate requirement for cellular UEs & $128$ Kbps \\
Rate requirement for D2D UEs & $256$ Kbps \\
Standard deviation of shadow fading: \\
 \hspace{5em} for relay-eNB links & $6$ dB \\
 \hspace{5em} for UE-relay links & $10$ dB \\
Noise power spectral density & $-174$ dBm/Hz \\
\hline
\end{tabular}
\end{table}

In our simulations, we assume $\omega = 1$, $\tilde{P}_{u_l,l}^{(n)}$ is set to $0$ dBm, and interference threshold is $-70~ \text{dBm}$ for all the RBs. The simulation parameters  are listed in Table \ref{tab:sim_param}. The simulation results are averaged over different realizations of  UE locations and channel gains.

\subsection{Results} \label{sec:numerical_results}

\subsubsection{Convergence} 

\begin{figure}
\centering
\includegraphics[scale=0.50]{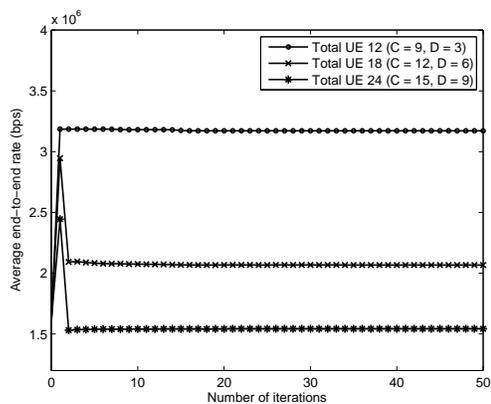}
\caption{Convergence behavior of the proposed algorithm with different number of UEs:  $D_{r,d} = 80~ \text{meter}$, $D_{d,d} = 140~ \text{meter}$.}
\label{fig:convergence_mp}
\end{figure}

In Fig. \ref{fig:convergence_mp}, we depict the convergence behavior of the proposed algorithm. In particular, we show the average achievable data
rate versus the number of iterations. The average achievable rate $R_{avg}$ for UEs is calculated as $R_{avg} =  \frac{ \displaystyle \sum_{u \in \lbrace\mathcal{C} \cup \mathcal{D} \rbrace} R_{u}^{ach}}{C+D}$ where $R_{u}^{ach}$ is the achievable data rate for UE $u$. Note that the higher
the number of users, the lower the average data rate.

\subsubsection{Performance of relay-aided D2D communication}

We compare the performance of the proposed scheme with the underlay D2D communication scheme presented in \cite{zul-d2d}. In this  \textit{reference scheme}, an RB allocated to CUE can be shared with at most one D2D-link. The D2D UEs share the same RB(s) (allocated to CUE using \textbf{Algorithm \ref{alg:mp_rec_alloc}}) and communicate directly between peers \textit{without} relay if the data rate requirements for both CUE and D2D UEs are satisfied; otherwise, the D2D UEs refrain from transmission on that particular time slot.

\begin{figure}
\centering
\includegraphics[scale=0.50]{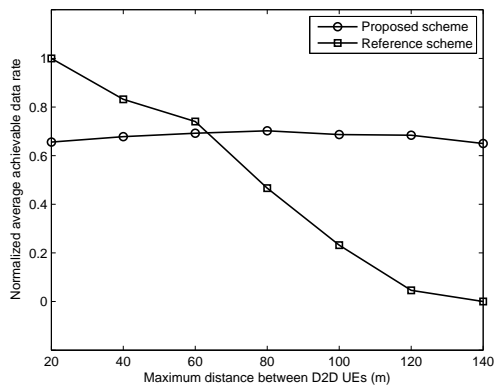}
\caption{Average achievable data rate for both the proposed and reference schemes with varying distance between D2D UEs: number of CUE,  $|\mathcal{C}| = 15$ and number of D2D pairs,  $|\mathcal{D}| = 9$ (i.e., $5$ CUE and $3$ D2D-pairs are assisted by each relay, and hence $|\mathcal{U}_l| = 8$ for each relay).  $D_{r,d}$ is considered $80~ \text{meter}$. } 
\label{fig:rate_05_03_mp}
\end{figure}

\begin{figure*}
\centering
\subfigure[]{\includegraphics[scale=0.50]{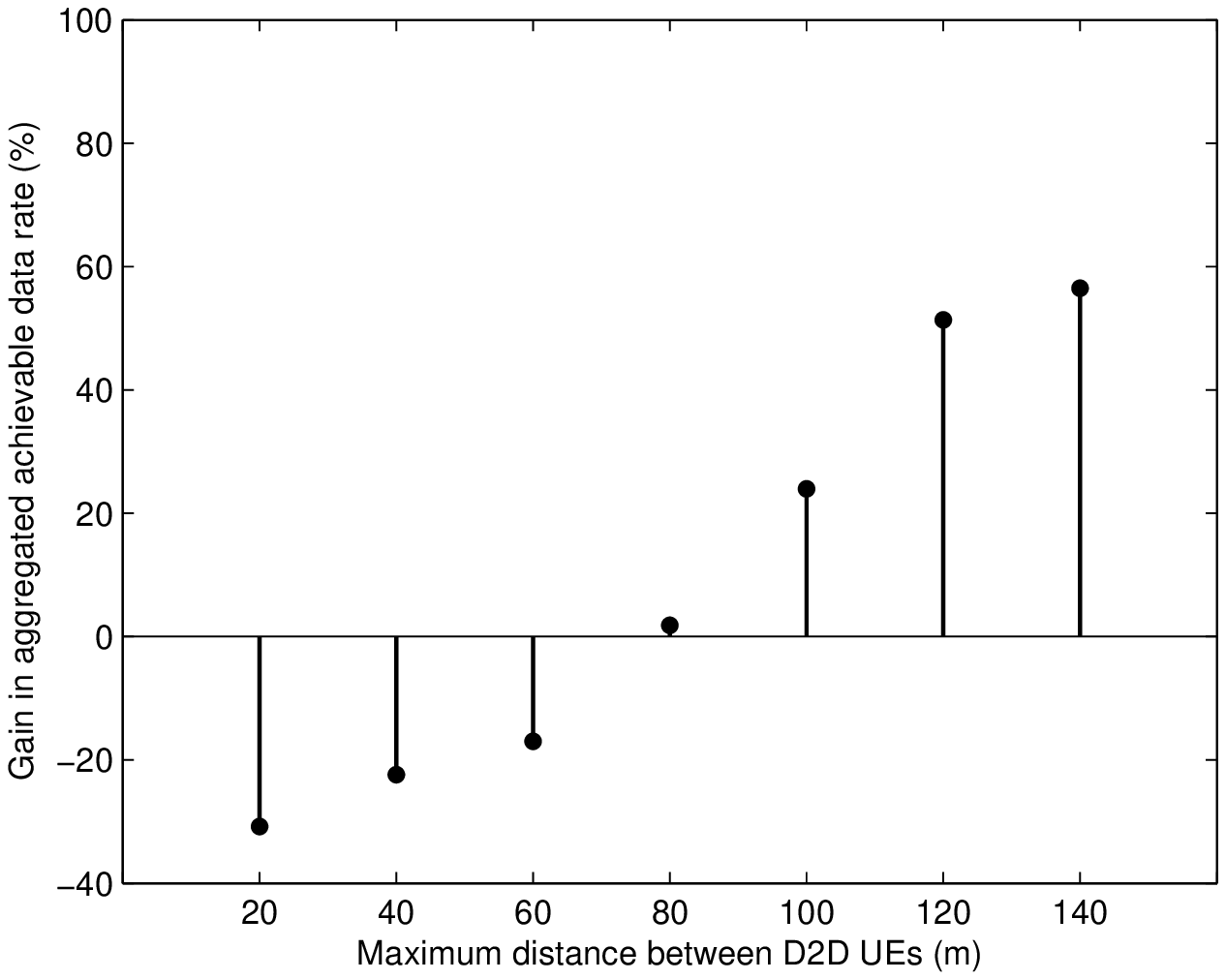}
\label{fig:gain_05_03_mp}}
\hfil 
\subfigure[]{\includegraphics[scale=0.50]{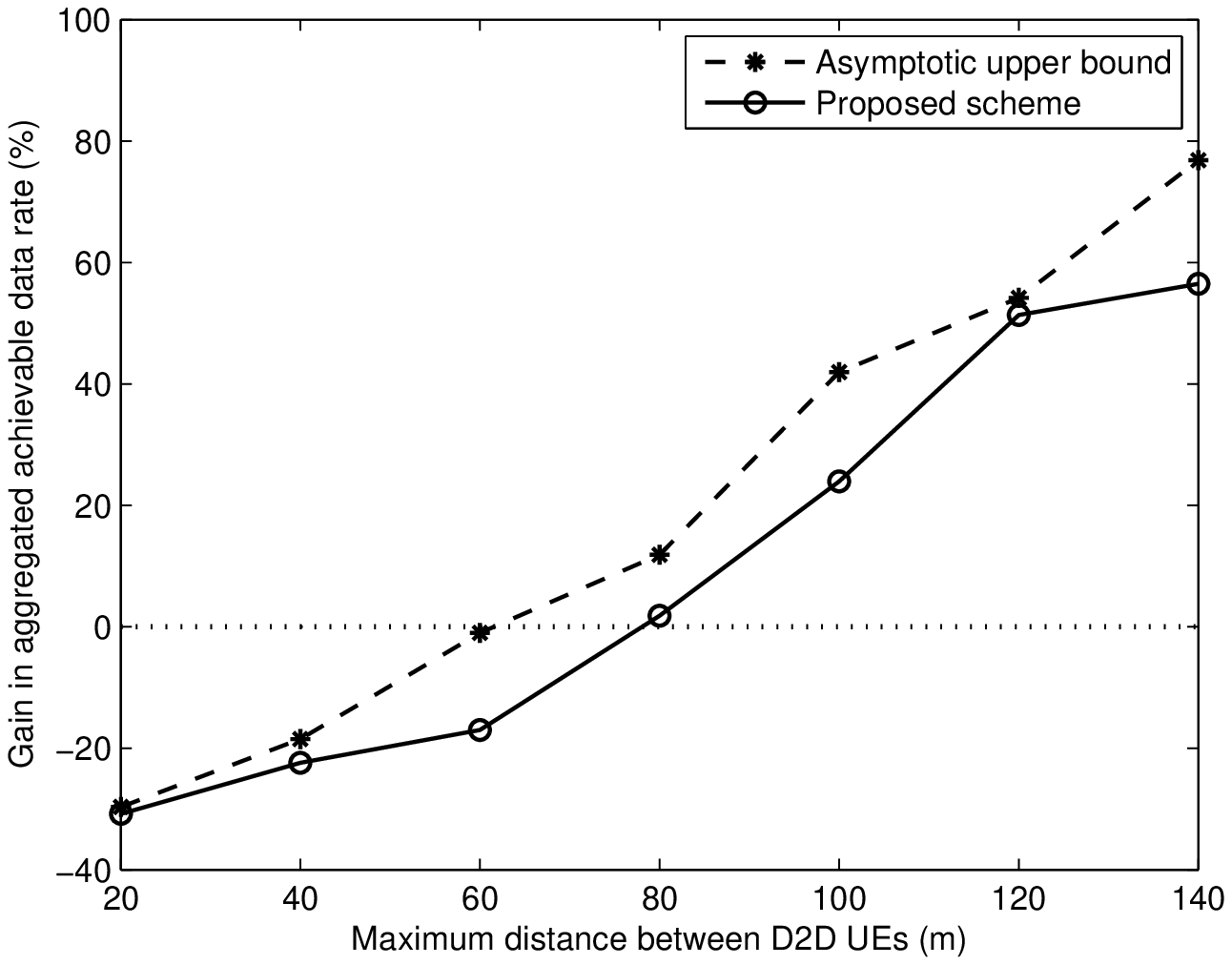}
\label{fig:gain_05_03_mp_compare}} 
\caption{(a)  Gain in aggregated achievable data rate and (b) Comparing gain with asymptotic upper bound using  the similar setup of Fig. \ref{fig:rate_05_03_mp}. There is a critical distance, beyond which relaying of D2D traffic provides significant performance gain.}
\label{fig:rate_gain_mp}
\end{figure*}

\begin{figure}[h t b]
\centering
\includegraphics[scale=0.50]{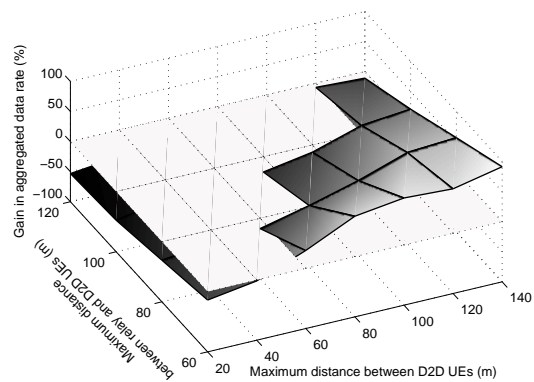}
\caption{Effect of relay distance on rate gain: $|\mathcal{C}| = 15$,  $|\mathcal{D}| = 9$. For every $D_{r,d}$, there is a distance threshold (i.e., upper side of the lightly shaded surface) beyond which relaying provides significant gain in terms of aggregated achievable rate.} 
\label{fig:rel_var_mp}
\end{figure}

\begin{figure}[h t b]
\centering
\includegraphics[scale=0.50]{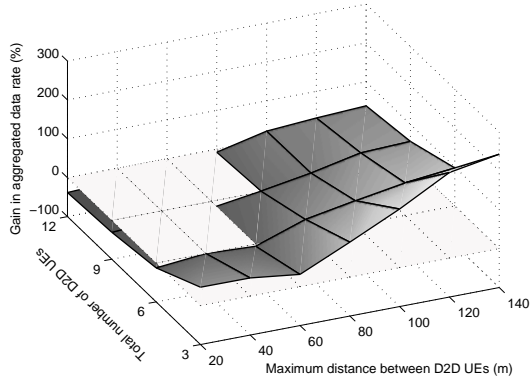}
\caption{Effect of number of D2D UEs on rate gain: $|\mathcal{C}| = 15$, $D_{r,d} = 80~ \text{meter}$. The upper position of lightly shaded surface illustrates the positive gain in terms of aggregated achievable rate.}
\label{fig:scal}
\end{figure}

\begin{figure}[h t b]
\centering
\includegraphics[scale=0.50]{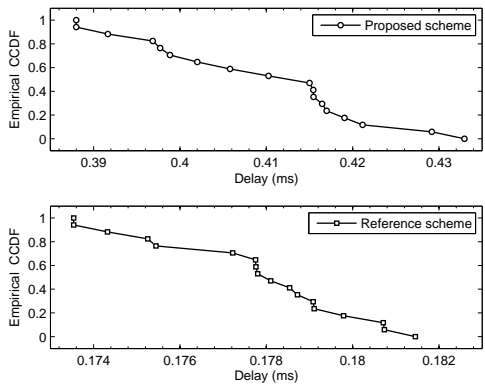}
\caption{End-to-end delay for the proposed and reference scheme where $|\mathcal{C}| = 15$,  $|\mathcal{D}| = 9$. We vary the distances $D_{r,d}$ and $D_{d,d}$ from $60$ to $140 ~\text{meter}$ with $5  ~\text{meter}$ interval. The decoding delay at a relay node is assumed to be $0.173~ \text{millisecond}$ (obtained from \cite{delay_imt}).}
\label{fig:delay}
\end{figure}

\begin{enumerate}[(i)]
\item \textit{Average achievable data rate vs. distance between D2D UEs:} The average achievable data rate of D2D UEs for both the proposed and reference schemes is illustrated in Fig. \ref{fig:rate_05_03_mp}. Although the reference scheme outperforms when the distance between D2D UEs is small (i.e., $d < 70~ \text{m}$), our proposed approach, which uses relays for D2D traffic, can greatly improve the data rate especially when the distance increases. This is due to the fact that when the distance increases, the performance of direct communication deteriorates due to increased signal attenuation. Besides, when the D2D UEs share resources with only one CUE, the spectrum may not be utilized efficiently, and therefore, the achievable rate decreases. As a result, the gap between the achievable rate with our proposed algorithm and that with the reference scheme becomes wider when the distance increases.

\item \textit{Gain in aggregated achievable data vs. varying distance between D2D UEs:} The gain in terms of aggregated achievable data rate is shown in Fig. \ref{fig:gain_05_03_mp}. We calculate the rate gain as follows: 
\begin{equation}
R_{gain} = \frac{R_{prop} - R_{ref}}{R_{ref}} \times 100 \%
\end{equation} 
where $R_{prop}$ and  $R_{ref}$ denote the aggregated data rate for the D2D UEs in the proposed scheme and the reference scheme, respectively. In Fig. \ref{fig:gain_05_03_mp_compare}, we compare the rate gain with the asymptotic upper bound\footnote{The asymptotic upper bound is obtained through relaxing the constraint that an RB is used by only one UE by using the time-sharing factor \cite{relax-con-1}. Thus $x_{u_l}^{(n)} \in (0,1]$ represents the sharing factor where each $x_{u_l}^{(n)}$ denotes the portion of time that RB $n$ is assigned to UE $u_l$ and satisfies the constraint $\displaystyle \sum_{u_l \in \mathcal{U}_l} x_{u_l}^{(n)} \leq 1, ~\forall n$. For details refer to \cite{d2d_our_paper}.}. The figures show that, compared to direct communication, with the increasing distance between D2D UEs, relaying provides considerable gain in terms of achievable data rate and hence spectrum utilization. In addition, our proposed distributed solution performs nearly close to the upper bound.

\item \textit{Effect of relay-UE distance and distance between D2D UEs on rate gain:} The performance gain in terms of the achievable aggregated data rate under different relay-D2D UE distance is shown in Fig. \ref{fig:rel_var_mp}. It is clear from the figure that, even for relatively large relay-D2D UE distances, e.g., $D_{r,d} \geq 80$ m, relaying D2D traffic provides considerable rate gain for distant D2D UEs.

\item \textit{Effect of number of D2D UEs and distance between D2D UEs on rate gain:} We vary the number of D2D UEs and plot the rate gain in Fig. \ref{fig:scal} to observe the performance of our proposed scheme in a dense network. The figure suggests that even in a moderately dense situation (e.g., $|\mathcal{C}| + |\mathcal{D}| = 15 + 12 = 27$) our proposed method provides a higher rate compared to direct communication between distant D2D UEs.

\item \textit{Impact of  relaying on delay:} In Fig. \ref{fig:delay}, we show results on the delay performance of the proposed relay-aided D2D communication approach. In particular, we observe the empirical complementary cumulative distribution function (CCDF)\footnote{The empirical CCDF of delay is defined as $\widehat{\mathfrak{D}}_\eta(\mathtt{t}) = \frac{1}{\eta} \displaystyle \sum_{i=1}^\eta \mathbb{I}_{[{\rm delay}_i > \mathtt{t}]}$ where $\eta$ is the total number of distance observations (e.g., UE-relay distance for the proposed scheme and the distance between D2D UEs for the reference scheme, respectively) used in the simulation, ${\rm delay}_i$ is the end-to-end delay at $i$-th distance observation, and $\mathtt{t}$  represents the $x$-axis values in Fig. \ref{fig:delay}. The indicator function $\mathbb{I}_{[\cdot]}$ outputs $1$ if the condition $[\cdot]$ is satisfied and $0$ otherwise.} for both the proposed scheme (which uses relay for D2D communication) and reference scheme (where D2D UEs communicate without relay). Note that in the reference scheme, the delay for one hop communication is given by $\mathfrak{D}_{\rm 1 hop} = \mathfrak{t}_{\rm schedule} + \mathfrak{t}_{\rm delivery}$. The variation in end-to-end delay is experienced due to variation in achievable data rate and propagation delay at different values of $D_{r,d}$ and $D_{d,d}$. From this figure it can be observed that the relay-aided D2D communication increases the end-to-end delay. However,  this increase (e.g., $0.431 - 0.189 = 0.242~  \text{msec}$) of delay would be acceptable for many D2D applications.


\end{enumerate}

\section{Conclusion} \label{sec:conclusion}

We have presented a comprehensive resource allocation
framework for relay-assisted D2D communication. Due to the $\mathcal{NP}$-hardness of original resource allocation problem, we have utilized the max-sum message passing strategy and presented a low-complexity distributed solution based on the message passing approach.  The convergence and optimality of the proposed scheme have been proved. The performance of the proposed method has been evaluated through simulations and we have observed that after a distance margin, relaying of D2D traffic improves system performance and provides a better data rate to the D2D UEs at the cost of a small increase in end-to-end delay.
In the context of D2D communication, most of the resource allocation problems are formulated under the assumption that the potential D2D UEs have already been discovered. However, to develop a complete D2D communication framework, this work can be extended considering D2D discovery along with resource allocation. In addition, due to time-varying and random nature of wireless channel, the link gain uncertainties for resource allocation in such relay-aided D2D communication is worth investigating.




\appendices
\numberwithin{equation}{section} 

\section{Required number of RB(s) for a given QoS requirement} 
\label{app:num_rb}

Let $\gamma_{u_l,l, 1}^{(n)}$ and $\bar{\gamma}_{u_l,l, 1}^{(n)}$ denote the instantaneous and average ${\rm SINR}$ for the UE $u_l$ over RB $n$.  In order to determine the required number of RB(s) for a given data rate requirement for any UE, we need to derive the probability distribution of $\frac{\gamma_{u_l,l, 1}^{(n)}}{\bar{\gamma}_{u_l,l, 1}^{(n)}}$  \cite{lee_qos}. Note that, the channel gain due to Rayleigh fading and log-normal shadowing can be approximated by a single log-normal distribution \cite{pdf_lognormal-1, pdf_lognormal}. In addition, the sum of random variables having log-normal distribution can be represented by a single log-normal distribution \cite{sum_lognormal}. Therefore, $\Gamma_{u_l,l, 1}^{(n)} = \frac{\gamma_{u_l,l, 1}^{(n)}}{\bar{\gamma}_{u_l,l, 1}^{(n)}}$ can be approximated by a log-normal random
variable whose mean and standard deviation can be calculated
as shown in \cite{pdf_lognormal}. Hence the average rate achieved by UE  $u_l$ over RB $n$ can be written as (\ref{eq:rate_pdf}) where $F_{\Gamma_{u_l,l, 1}^{(n)}}(\vartheta)$ and $f_{\Gamma_{u_l,l, 1}^{(n)}}(\vartheta) $ are the probability density function and probability distribution function of $\Gamma_{u_l,l, 1}^{(n)}$, respectively.

\begin{figure*}[!t]
\normalsize

\begin{equation} \label{eq:rate_pdf}
\bar{r}_{u_l, l}^{(n)} = \frac{1}{2} B_{RB} \hspace{-0.3em} \bigints_{0}^{\infty} \hspace{-1em}\log_2 \left( 1 +  P_{u_l, l}^{(n)} \Gamma_{u_l,l, 1}^{(n)} \bar{\gamma}_{u_l,l, 1}^{(n)}\right) \left[ \prod_{\substack{j \in \mathcal{U}_l, \\ j \neq u_l}}  F_{\Gamma_{j,l, 1}^{(n)}}(\vartheta) \right] f_{\Gamma_{u_l,l, 1}^{(n)}}(\vartheta) d\vartheta.
\end{equation}

\hrulefill
\vspace*{4pt}
\end{figure*}

Now, let $\underline{R}_{u_l, l}$ be the minimum rate achieved by UE $u_l$. In order to maintain the data rate requirement, we can derive the following inequality\footnote{Similar to \cite{lee_qos}, we assume that the long-term channel gains on different RBs are same, and hence, the average rates achieved by a particular UE on different RBs are the same.}: 
\begin{equation}
Q_{u_l} \leq \kappa_{u_l} \leq \underline{R}_{u_l, l}\left( |\mathcal{U}_l| \right)
\end{equation}
where by $ \underline{R}_{u_l, l}\left( |\mathcal{U}_l| \right)$ we explicitly describe the dependence of the minimum achievable rate $\underline{R}_{u_l, l}$  on the number of  UEs $|\mathcal{U}_l|$. Therefore, the minimum number of required RBs is given by
\begin{equation}
\kappa_{u_l} \geq \left \lceil \frac{Q_{u_l}}{\underline{R}_{u_l, l}\left( |\mathcal{U}_l| \right)} \right \rceil.
\end{equation}

\section{Proof of Theorem \ref{theorem:optimal-proof-mp}} 
\label{app:optimal-mp}

\begin{figure*}[!t]
\normalsize

\begin{flalign}
\boldsymbol{\mathbb{L}}_l   =  \nonumber  
& \sum_{u_l \in \mathcal{U}_l} \sum_{n = 1}^{N} \tfrac{1}{2} x_{u_l}^{(n)} R_{u_l}^{(n)} + \sum_{n = 1}^{N}  \ddot{a}_n  \left( 1 - \sum_{u_l \in \mathcal{U}_l} x_{u_l}^{(n)}  \right) 
+ \sum_{u_l \in \mathcal{U}_l}  \ddot{b}_{u_l} \left(P_{u_l}^{max} - \sum_{n = 1}^{N} x_{u_l}^{(n)} P_{u_l, l}^{(n)}  \right) \\
& + \ddot{c}_l \left( P_l^{max} - \sum_{u_l \in \mathcal{U}_l } \sum_{n =1}^N \tfrac{\gamma_{u_l, l, 1}^{(n)}}{\gamma_{l, u_l, 2}^{(n)}} x_{u_l}^{(n)} P_{u_l, l}^{(n)}  \right) 
+ \sum_{n = 1}^{N}  \ddot{d}_n  \left(I_{th, 1}^{(n)} - \sum_{u_l \in \mathcal{U}_l } x_{u_l}^{(n)} P_{u_l, l}^{(n)} g_{{u_l^*}, l, 1}^{(n)}  \right) \nonumber \\
&+ \sum_{n = 1}^{N}  \ddot{e}_n  \left(I_{th, 2}^{(n)}  - \sum_{u_l \in \mathcal{U}_l } \tfrac{\gamma_{u_l, l, 1}^{(n)}}{\gamma_{l, u_l, 2}^{(n)}} x_{u_l}^{(n)} P_{u_l, l}^{(n)} g_{l, {u_l^*}, 2}^{(n)} \right) 
+ \sum_{u_l \in \mathcal{U}_l }  \ddot{f}_{u_l} \left( \sum_{n=1}^N \tfrac{1}{2}  x_{u_l}^{(n)} R_{u_l}^{(n)} - Q_{u_l} \right).\label{eq:lagrange-1-mp} 
\end{flalign}

\hrulefill
\vspace*{4pt}
\end{figure*}

Let us rearrange the Lagrangian of $\mathbf{P2}$  defined by (\ref{eq:lagrange-1-mp}) as follows:
\begin{flalign}
\boldsymbol{\mathbb{L}}_l   =  \nonumber  
& \sum_{u_l \in \mathcal{U}_l} \sum_{n = 1}^{N} \tfrac{1}{2} x_{u_l}^{(n)} R_{u_l}^{(n)} - \sum_{n = 1}^{N}  \ddot{a}_n \sum_{u_l \in \mathcal{U}_l} x_{u_l}^{(n)}  \nonumber \\
&- \sum_{u_l \in \mathcal{U}_l}  \ddot{b}_{u_l} \sum_{n = 1}^{N} x_{u_l}^{(n)} P_{u_l, l}^{(n)}  - \ddot{c}_l \sum_{u_l \in \mathcal{U}_l } \sum_{n =1}^N \tfrac{\gamma_{u_l, l, 1}^{(n)}}{\gamma_{l, u_l, 2}^{(n)}} x_{u_l}^{(n)} P_{u_l, l}^{(n)}  \nonumber \\
&- \sum_{n = 1}^{N}  \ddot{d}_n  \sum_{u_l \in \mathcal{U}_l } x_{u_l}^{(n)} P_{u_l, l}^{(n)} g_{{u_l^*}, l, 1}^{(n)} \nonumber \\
&- \sum_{n = 1}^{N}  \ddot{e}_n  \sum_{u_l \in \mathcal{U}_l } \tfrac{\gamma_{u_l, l, 1}^{(n)}}{\gamma_{l, u_l, 2}^{(n)}} x_{u_l}^{(n)} P_{u_l, l}^{(n)} g_{l, {u_l^*}, 2}^{(n)}  \nonumber \\
&- \sum_{u_l \in \mathcal{U}_l }  \ddot{f}_{u_l}   \sum_{n=1}^N \tfrac{1}{2}  x_{u_l}^{(n)} R_{u_l}^{(n)} + \widetilde{O}.\label{eq:lagrange-1} 
\end{flalign}
where $\widetilde{O}$ denote the leftover terms involving Lagrange multipliers, i.e., $\boldsymbol{\ddot{a}}, \boldsymbol{\ddot{b}}, \boldsymbol{\ddot{c}}, \boldsymbol{\ddot{d}}, \boldsymbol{\ddot{e}}, \boldsymbol{\ddot{f}}$. From above we can derive the following lemma:

\begin{appxlem} \label{lemma:slackness}
The slackness conditions for $\mathbf{P2}$ are 
\begin{equation} \label{eq:slackness-mp}
\widehat{R}_{u_l}^{(n)} - \ddot{\lambda}_{u_l}^{{(n)}^*} = 
\underset{1 \leq j \leq N}{ \operatorname \max} \left(   \widehat{R}_{u_l}^{(j)} - \ddot{\lambda}_{u_l}^{{(j)}^*}  \right) 
\end{equation}
where $\ddot{\lambda}_{u_l}^{(n)}$ involves the terms with Lagrange multipliers for $\forall u_l, n$.
\end{appxlem}
\begin{IEEEproof}
By Weierstrass' theorem (Appendix A.2, Proposition A.8 in \cite{nonlinear_book_mp}) the dual function can be calculated by (\ref{eq:conv-mp2}).

\begin{figure*}[!t]
\normalsize

\begin{flalign} 
\mathscr{D}_l &= \underset{\boldsymbol{x_l}}{ \operatorname \inf} ~ \boldsymbol{\mathbb{L}}_l  \nonumber \\
&= \underset{\boldsymbol{x_l}}{ \operatorname \inf} \nonumber  
\sum_{u_l \in \mathcal{U}_l} \left( \sum_{n = 1}^{N} \left(  \tfrac{1}{2}  R_{u_l}^{(n)} - \ddot{a}_n - \ddot{b}_{u_l} P_{u_l, l}^{(n)}  - \ddot{c}_l  \tfrac{\gamma_{u_l, l, 1}^{(n)}}{\gamma_{l, u_l, 2}^{(n)}}  P_{u_l, l}^{(n)} - \ddot{d}_n   P_{u_l, l}^{(n)} g_{{u_l^*}, l, 1}^{(n)} 
- \ddot{e}_n   \tfrac{\gamma_{u_l, l, 1}^{(n)}}{\gamma_{l, u_l, 2}^{(n)}}  P_{u_l, l}^{(n)} g_{l, {u_l^*}, 2}^{(n)} +  \ddot{f}_{u_l}   \tfrac{1}{2}   R_{u_l}^{(n)}  \right) x_{u_l}^{(n)} \right) + \widetilde{O} \nonumber \\
&= \sum_{u_l \in \mathcal{U}_l} \left( \sum_{n = 1}^{N} \underset{\boldsymbol{x_l}}{ \operatorname \inf} \left(  \tfrac{1}{2}  R_{u_l}^{(n)} (1 +  \ddot{f}_{u_l}) - \ddot{\lambda}_{u_l}^{(n)}  \right) x_{u_l}^{(n)} \right) + \widetilde{O} 
=\sum_{u_l \in \mathcal{U}_l} \underset{1 \leq n \leq N}{ \operatorname \max} \left(   \widehat{R}_{u_l}^{(n)} - \ddot{\lambda}_{u_l}^{(n)}  \right) \kappa_{u_l} + \widetilde{O}. 
 \label{eq:conv-mp2} 
\end{flalign}

\hrulefill
\vspace*{4pt}
\end{figure*}

Therefore, if $\mathbf{P2}$ has an optimal solution, its dual has an optimal solution, i.e., 
\begin{equation}
{\mathscr{D}_l}^* = \sum_{u_l \in \mathcal{U}_l} \sum_{n = 1}^{N} \widehat{R}_{u_l}^{(n)} {x_{u_l}^{(n)}}^*.
\end{equation}
Hence, 
\begin{align} \label{eq:mp_dual_opt}
\sum_{u_l \in \mathcal{U}_l} \underset{1 \leq n \leq N}{ \operatorname \max} \left(   \widehat{R}_{u_l}^{(n)} - \ddot{\lambda}_{u_l}^{{(n)}^*}  \right) \kappa_{u_l} + \widetilde{O} = \sum_{u_l \in \mathcal{U}_l} \sum_{n = 1}^{N} \widehat{R}_{u_l}^{(n)} {x_{u_l}^{(n)}}^*.
\end{align}
Since ${\boldsymbol{x_l}}^*$ is an optimal allocation, from (\ref{eq:mp_dual_opt}) we obtain
\begin{align} \label{eq:mp_dual_opt2}
\sum_{u_l \in \mathcal{U}_l} \underset{1 \leq n \leq N}{ \operatorname \max} & \left(   \widehat{R}_{u_l}^{(n)} - \ddot{\lambda}_{u_l}^{{(n)}^*}  \right) \kappa_{u_l} \nonumber \\ 
&= \sum_{u_l \in \mathcal{U}_l} \sum_{n = 1}^{N} \left( \widehat{R}_{u_l}^{(n)} - \ddot{\lambda}_{u_l}^{{(n)}^*} \right) {x_{u_l}^{(n)}}^*.
\end{align}
In addition, since $\displaystyle \sum_{n=1}^{N} x_{u_l}^{(n)} = \kappa_{u_l}$, (\ref{eq:mp_dual_opt2}) becomes
\begin{equation}
\sum_{u_l \in \mathcal{U}_l} \sum_{n = 1}^{N} \left(  \widehat{R}_{u_l}^{(n)}  - \ddot{\lambda}_{u_l}^{{(n)}^*}  - \underset{1 \leq n \leq N} { \operatorname \max} \left(   \widehat{R}_{u_l}^{(n)} - \ddot{\lambda}_{u_l}^{{(n)}^*}  \right) \right) {x_{u_l}^{(n)}}^* = 0.
\end{equation}
Now, if ${x_{u_l}^{(n)}}^*>0$, we have $$\widehat{R}_{u_l}^{(n)} - \ddot{\lambda}_{u_l}^{{(n)}^*} = 
\underset{1 \leq j \leq N}{ \operatorname \max} \left(   \widehat{R}_{u_l}^{(j)} - \ddot{\lambda}_{u_l}^{{(j)}^*}  \right). $$
\end{IEEEproof}

From (\ref{eq:rb_alloc_mp}), at each iteration, each UE $u_l$ can distinguish between two different subsets of RBs by sorting the marginals in an increasing order. Let us define  the first subset $\dot{\mathcal{N}}_{u_l} \in \mathcal{N}$ given by the first $\kappa_{u_l} \leq N$ RBs in the ordered list of marginals where the second subset $\ddot{\mathcal{N}}_{u_l} \in \mathcal{N}$ is given by the last $N - \kappa_{u_l}$ of the list. Accordingly, we can have the following lemma:
\begin{appxlem} \label{lemma:con-mp-2}
At convergence, $\widehat{R}_{u_l}^{(\dot{n})} + \tilde{\psi}_{u_l,l}^{(\dot{n})} < \widehat{R}_{u_l}^{(\ddot{n})} + \tilde{\psi}_{u_l,l}^{(\ddot{n})}$ ~ for $\forall u_l, \dot{n} \in \dot{\mathcal{N}}_{u_l}, \ddot{n} \in \ddot{\mathcal{N}}_{u_l}$.
\end{appxlem}
\begin{IEEEproof}
See \cite{min-sum-mp}. 
\end{IEEEproof}

From  \textbf{Lemma \ref{lemma:slackness}} and \textbf{\ref{lemma:con-mp-2}}, it can be noted that, the inequality $\widehat{R}_{u_l}^{(\dot{n})} - \ddot{\lambda}_{u_l}^{{(\dot{n})}^*} < \widehat{R}_{u_l}^{(\ddot{n})} - \ddot{\lambda}_{u_l}^{{(\ddot{n})}^*}$ implies the slackness condition (\ref{eq:slackness-mp}) by imposing $\ddot{\lambda}_{u_l}^{{(\dot{n})}^*} = - \tilde{\psi}_{u_l,l}^{(\dot{n})}$ and $\ddot{\lambda}_{u_l}^{{(\ddot{n})}^*} = - \tilde{\psi}_{u_l,l}^{(\ddot{n})}$; hence, the proof of \textbf{Theorem \ref{theorem:optimal-proof-mp}} follows.

\section{Proof of Theorem \ref{theorem:converge}} 
\label{app:converge-mp}

From \cite{large-rb-dual} and Proposition 4 of \cite{mp-convergence-proof},  there must exist a non-overlapping binary valued feasible allocation even after relaxation when the number of RBs tends to infinity. Since in our problem the number of RBs is sufficiently large, the messages converge to a fixed point and we can conclude that the LP relaxation of $\mathbf{P2}$, i.e., $x_{u_l}^{(n)} \in (0,1]$ achieves the same optimal objective value. Thus, directly following the theorem of integer programming duality (i.e., if the primal problem has an optimal solution, then the dual also has an optimal one) for any finite $N$, the optimal objective value of $\mathscr{D}_l$ lies between $\mathbf{P2}$ and its LP relaxation.

%

\bibliographystyle{IEEEtran}



\begin{IEEEbiography} [{\includegraphics[width=1in,height=1.25in,clip,keepaspectratio]{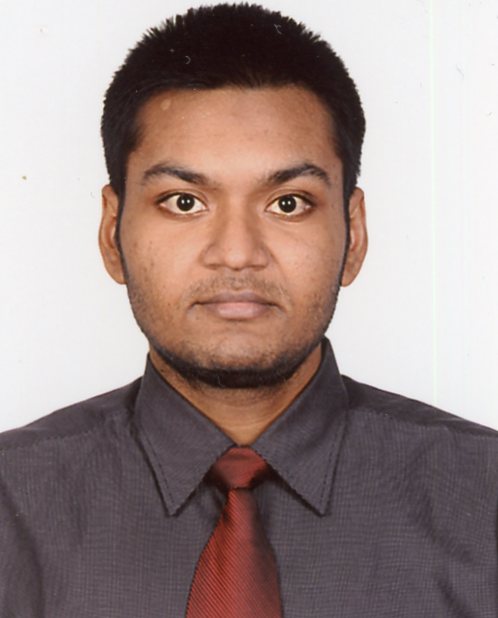}}]
{Monowar Hasan} (S'13)  received his B.Sc. degree in Computer Science and Engineering from Bangladesh University of Engineering and Technology (BUET), Dhaka, in 2012. He is currently working toward his M.Sc. degree in the Department of Electrical and Computer Engineering at the University of Manitoba, Winnipeg, Canada. He has been awarded the University of Manitoba Graduate Fellowship. His current research interests include internet of things, software defined networks and resource allocation in mobile cloud computing. He serves as a reviewer for several major IEEE journals and conferences.

\end{IEEEbiography}

\begin{IEEEbiography} [{\includegraphics[width=1in,height=1.25in,clip,keepaspectratio]{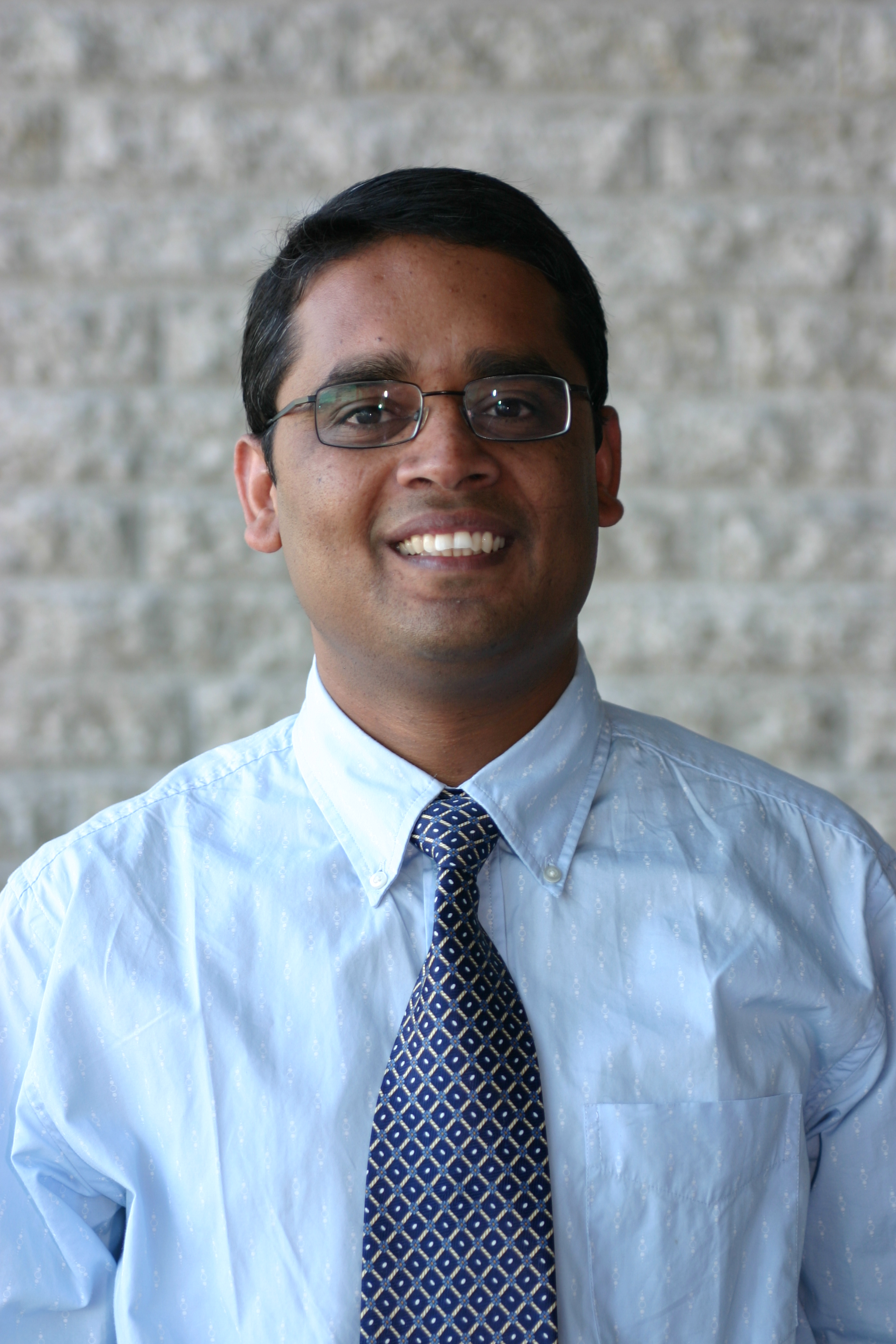}}]
{Ekram Hossain} (S'98-M'01-SM'06)  
is a Professor (since March 2010) in the Department of Electrical and Computer Engineering at University of Manitoba, Winnipeg, Canada. He received his Ph.D. in Electrical Engineering from University of Victoria, Canada, in 2001. Dr. Hossain's current research interests include design, analysis, and optimization of wireless/mobile communications networks, cognitive radio systems, and network economics.  He has authored/edited several books in these areas (http://home.cc.umanitoba.ca/$\sim$hossaina). His research has been widely cited in the literature (more than 7000 citations in
Google Scholar with an h-index of 42 until January 2014). Dr. Hossain  serves as the Editor-in-Chief for the {\em IEEE Communications Surveys and Tutorials}  and an Editor for {\em IEEE Journal on Selected Areas in Communications - Cognitive Radio Series} and {\em IEEE Wireless Communications}.  Also, he is a member of the IEEE Press Editorial Board. Previously, he served as the Area Editor for the {\em IEEE Transactions on Wireless Communications} in the area of  ``Resource Management and Multiple Access'' from 2009-2011 and an Editor for the IEEE Transactions on Mobile Computing from 2007-2012. He is also a member of the IEEE Press Editorial Board. Dr. Hossain has won several research awards including the University of Manitoba Merit Award in 2010 (for Research and Scholarly Activities), the 2011 IEEE Communications Society Fred Ellersick Prize Paper Award, and the IEEE Wireless Communications and Networking Conference 2012 (WCNC'12) Best Paper Award. He is a Distinguished Lecturer of the IEEE Communications Society (2012-2015). Dr. Hossain is a registered Professional Engineer in the province of Manitoba, Canada. 
\end{IEEEbiography}

\end{document}